\documentclass[10pt,journal,compsoc]{IEEEtran}


\usepackage[normalem]{ulem}
\usepackage{makecell}
\usepackage{pifont}
\usepackage{rotating, multirow}
\usepackage{tikz}
\useunder{\uline}{\ul}{}
\usepackage{subcaption}
\usepackage{xspace}
\usepackage{enumitem}
\usepackage[]{algorithm2e}
\usepackage{colortbl}
\usepackage{fontawesome5}
\usepackage{tabularx}
\usepackage{csquotes}
\usepackage{textcomp}
\usepackage{color}
\usepackage{comment}
\usepackage{hyperref}
\usepackage{booktabs}
\usepackage{blindtext}
\usepackage{amssymb}
\usepackage[T1]{fontenc}
\usepackage{cite} 

\newif\ifreview

\newcommand\reviewcomment[3]{
    \ifreview
    \textcolor{#1}{#2 - #3}
    \fi
}

\definecolor{DarkCerulean}{rgb}{0.03, 0.27, 0.49}

\newcommand\af[1]{\reviewcomment{orange}{#1}{AF}}

\newcommand\kh[1]{\reviewcomment{gray}{#1}{KH}}

\newcommand{\signal}[1] {\textit{#1}}

\newcommand{\rot}[2]{\rotatebox{#1}{#2}}

\definecolor{marketentrycolor}{HTML}{bd6b19}
\definecolor{developernamecolor}{HTML}{507dbc}
\definecolor{developerwebsitecolor}{HTML}{345511}
\definecolor{developeremailcolor}{HTML}{adaabf}
\definecolor{privacypolicycolor}{HTML}{7a3b69}
\definecolor{appnamecolor}{HTML}{fb1363}
\definecolor{packagenamecolor}{HTML}{000000}
\definecolor{certcolor}{HTML}{000000}
\definecolor{Gray}{gray}{0.9}
\newcommand{\tikzcircle}[2][red,fill=red]{\tikz[baseline=-0.5ex]\draw[#1,radius=#2] (0,0) circle ;}\definecolor{barcolor}{HTML}{507dbc}

\newcommand{\y}{\ding{51}}
\newcommand{\n}{}

\newcommand{\notcollectedsignal}{---}

\newcommand{\finished}[1]{#1}

\providecommand{\ie}{\textit{i.e.,}\xspace}
\providecommand{\eg}{\textit{e.g.,}\xspace}

\providecommand{\etal}{\textit{et al.}\xspace}

\newtheorem{definition}{Definition}

\newcommand{\parax}[1]{\noindent\textbf{#1.}}
\newcommand{\rquestion}[1]{\textit{``#1''}}

\renewcommand{\texttt}[1]{\begingroup
 \ttfamily
 \begingroup\lccode`~=`/\lowercase{\endgroup\def~}{/\discretionary{}{}{}}\begingroup\lccode`~=`[\lowercase{\endgroup\def~}{[\discretionary{}{}{}}\begingroup\lccode`~=`.\lowercase{\endgroup\def~}{.\discretionary{}{}{}}\catcode`/=\active\catcode`[=\active\catcode`.=\active
 \scantokens{#1\noexpand}\endgroup
}

\usepackage{fp}
\newcommand{\maxnum}{100.00}
\newlength{\maxlen}
\newcommand{\databar}[2][barcolor]{\settowidth{\maxlen}{\maxnum}\addtolength{\maxlen}{\tabcolsep}\FPeval\result{round(#2/\maxnum:4)}\FPeval\toprint{clip(round(#2:1))}\rlap{\color{#1!40}\hspace*{-.5\tabcolsep}\rule[-.05\ht\strutbox]{\result\maxlen}{.95\ht\strutbox}}\makebox[\dimexpr\maxlen-\tabcolsep][r]{\toprint\%}}
\newcommand{\databarfull}[2][barcolor]{\settowidth{\maxlen}{\maxnum}\addtolength{\maxlen}{\tabcolsep}\FPeval\result{round(#2/\maxnum:4)}\FPeval\toprint{#2}\rlap{\color{#1!40}\hspace*{-.5\tabcolsep}\rule[-.05\ht\strutbox]{\result\maxlen}{.95\ht\strutbox}}\makebox[\dimexpr\maxlen-\tabcolsep][r]{\toprint\%}}

\newcolumntype{L}[1]{>{\raggedright\let\newline\\\arraybackslash\hspace{0pt}}m{#1}}
\newcolumntype{C}[1]{>{\centering\let\newline\\\arraybackslash\hspace{0pt}}m{#1}}
\newcolumntype{R}[1]{>{\raggedleft\let\newline\\\arraybackslash\hspace{0pt}}m{#1}}

\newcommand{\www}{Proceedings of the International Conference on World Wide Web (WWW)}
\newcommand{\imc}{Proceedings of the Internet Measurement Conference (IMC)}

\newcommand{\msr}{Proceedings of the International Conference on Mining Software Repositories (MSR)}

\newcommand{\ndss}{Proceedings of the Network and Distributed System Security Symposium (NDSS)}

\newcommand{\icse}{Proceedings of the International Conference on Software Engineering}
\newcommand{\dimva}{Proceedings of the International conference on detection of intrusions and malware, and vulnerability assessment (DIMVA)}

\newcommand{\sigsac}{Proceedings of the ACM SIGSAC Conference on Computer and Communications Security}
\newcommand{\esoric}{Proceedings of the European Symposium on Research in Computer Security (ESORICS)}

\newcommand{\usenix}{Proceedings of the USENIX Security Symposium}
\newcommand{\conpro}{Workshop on Technology and Consumer Protection (ConPro)}

\newcommand{\codaspy}{Proceedings of the ACM Conference on Data and Application Security and Privacy (CODASPY)}
\newcommand{\oakland}{IEEE Symposium on Security and Privacy (SP)}

\newcommand{\sigmobile}{Proceedings of the SIGMOBILE Mobile Computing and Communications Review}

\newcommand{\spsm}{Proceedings of the Second ACM Workshop on Security and Privacy in Smartphones and Mobile Devices}
\newcommand{\issta}{Proceedings of the International Symposium on Software Testing and Analysis}

\newcommand{\frommarket}{\faShoppingCart}
\newcommand{\frommanifest}{\faFileCode[regular]}
\newcommand{\fromcertificate}{\faAward}

\newcommand{\frommarketandcert}{\frommarket{} + \fromcertificate{}}
\newcommand{\frommarketandmanifest}{\frommarket{} + \frommanifest{}}

\newcommand{\rqone}{How available and volatile are attribution signals on Android markets?}
\newcommand{\rqtwo}{How consistent are attribution signals within apps, within markets and across them?}
\newcommand{\rqthree}{How is the Google Play Store affected by the lack of signal availability and consistency?}

\newcommand{\rqoneref}{\hyperlink{rqone}{\textbf{RQ1}}}
\newcommand{\rqtworef}{\hyperlink{rqtwo}{\textbf{RQ2}}}
\newcommand{\rqthreeref}{\hyperlink{rqthree}{\textbf{RQ3}}} 
\makeatletter
\def\@rothead[#1]#2{\thead{\\[-.65\normalbaselineskip]
  \turn{\cellrotangle}\thead[#1]{#2}\endturn}}
\makeatother

\reviewtrue

\begin{document}
    \title{Mixed Signals: Analyzing Software Attribution Challenges in the Android Ecosystem}

    \author{Kaspar~Hageman, \'Alvaro~Feal, Julien~Gamba, Aniketh~Girish, Jakob~Bleier, Martina~Lindorfer, Juan~Tapiador, Narseo~Vallina-Rodriguez}

  \IEEEtitleabstractindextext{\begin{abstract}
The ability to identify the author responsible for a given software
object is critical for many research studies 
and for enhancing software transparency and accountability. However,
as opposed to other application markets like iOS, attribution in the
Android ecosystem is known to be hard. 
Prior research has leveraged market metadata and
signing certificates to identify software authors without questioning
the validity and accuracy of these attribution signals.
However, Android app authors can, either intentionally or by mistake,
hide their true identity due to: 
(1) the lack of policy enforcement by markets to ensure the
accuracy and correctness of the information disclosed
by developers in their market profiles during the app release process, and 
(2) the use of self-signed certificates for signing apps instead of
certificates issued by trusted CAs. 
 
In this paper, we perform the first empirical analysis
of the availability, volatility and overall aptness of publicly available
metadata for author attribution in Android app markets. 
To that end, we analyze a dataset of over
2.5 million market entries and apps extracted from five Android markets
for over two years. Our results show that widely used attribution signals are 
often missing from market profiles and that they change over time. We also 
invalidate the general belief about the
validity of signing certificates 
for author attribution. For instance, we find
that apps from different authors share signing certificates
due to the proliferation of app building frameworks and software factories.
Finally, we introduce the concept of attribution graph and 
we apply it to evaluate the validity of existing attribution signals
on the Google Play Store. Our results confirm that  
the lack of control over publicly available signals can 
confuse the attribution process.
\end{abstract}

}

    \maketitle

    \IEEEdisplaynontitleabstractindextext
    \IEEEpeerreviewmaketitle

    \IEEEraisesectionheading{\section{Introduction}\label{sec:introduction}}

Software \textit{attribution} is the process of matching a piece of software
to its author. This concept has gained attention in the research 
community as it is critical for software analysis, 
platform measurements, security and threat analysis, transparency, and 
regulatory enforcement~\cite{lindorfer2014andrubis,
alrabaee2016feasibility,platon2020pup,gdpr_rights, explorativemobile2017, 
wang2018beyond, wangwww2019, ali2017marketcompare, Zhong2013longtail,
viennot2014measurement, Holzer2011devperps, Heureuse2012whatsanapp}.
Every software platform implements different attribution mechanisms.
Windows relies on a Public Key Infrastructure (PKI) that provides
authenticity guarantees about the organization offering the software  
through the use of X.509 certificates issued by trusted Certificate 
Authorities (CAs)~\cite{kaliski1998pkcs, kim2017certified,
vandersloot2016towards}.
In the case of iOS, apps must be signed with a developer certificate issued by 
Apple~\cite{apple-dev-enroll,apple-identity-verification}.
These certificates are part of Apple's developer program, which
involves the verification of developers' legal
identity~\cite{apple-dev-enroll,apple-identity-verification}.

Android implements laxer attribution schemes regardless of the app market.
During the development and publication process of apps, 
developers can disclose attribution data both during the app signing
process and on their market profile
(\eg{} developer name, email and website)~\cite{playStoreRulesMetadata}.
This information is \emph{self-declared by the developer} and is not
endorsed nor validated by a trusted authority.
Even the cryptographic signing certificates can be
self-signed~\cite{google-play-recommended-certificate-lifetime}.
While other software platforms also distribute software under potentially unverified,
self-declared attribution data, their PKI ensures some form of
control by the platform operator which is absent in the Android ecosystem.
To complicate things further, the diversity of
publication policies across Android app markets 
translates into a lack of a robust
Android-wide attribution mechanisms~\cite{lindorfer2014andradar, wang2018beyond}.
This state of affairs impedes 
external actors, such as researchers and regulators (and, possibly,
the market operators themselves) from automatically studying developer practices, 
enhancing software accountability, or effectively detecting harmful, cloned
and deceptive apps~\cite{lindorfer2014andrubis, okoyomon2019ridiculousness,
crussell2012attack,kim2019romadroid, 
wang2015wukong,chen2014achieving}.
As a result, end users are potential victims of 
impersonation attacks, such as repackaged malware~\cite{crussell2012attack} or phishing
attacks~\cite{droideagle2015suna}, which may also have a negative impact
on the revenue streams and reputation of legitimate developers. 

Prior research has relied on self-declared data, 
such as the app certificate~\cite{crussell2012attack,
Empiricalinstallation2012, vendorcustom2013, viennot2014measurement,
lindorfer2014andradar, findingmalice2015, gonzalez2018authorship,
kalgutkar2018android, oltrogge2018rise, wang2018beyond, li2019rebooting,
androzoo2020, gamba2020analysis, platon2020pup, sebastian2020towards},
app name~\cite{ali2017marketcompare, xucodeattribution2019,
platon2020pup,sebastian2020towards}, the package
name~\cite{lindorfer2014andradar, oltrogge2018rise, Phishinginandroid2018}, 
or market metadata~\cite{thomasimc2013, viennot2014measurement, ali2017marketcompare,
explorativemobile2017, Phishinginandroid2018, wangwww2019,
xucodeattribution2019, incentivized2020farooqi, platon2020pup,
sebastian2020towards} for author attribution. 
In some cases, authors combined multiple signals
hoping to increase their strength. However, none of
these approaches have been sufficiently validated and the general
attribution problem remains poorly understood by the community.
This paper fills this knowledge gap in the Android
ecosystem by assessing 
the validity of publicly available 
attribution signals, including app market metadata
and signing certificates. 
Specifically, we answer the following research questions:

\parax{\hypertarget{rqone}{RQ1}}\xspace \rqone

\parax{\hypertarget{rqtwo}{RQ2}}\xspace \rqtwo

\parax{\hypertarget{rqthree}{RQ3}}\xspace \rqthree

To address these questions, we follow an empirical approach using large-scale, real-world data,
and discuss the implications of these measurements.
First, we review the Android app signing process and
the policies defined by different markets to report app authors (\S\ref{sec:background}). Our findings inform the definition of a set of
attribution signals (\S\ref{sec:research_questions}) that we
use to conduct a large-scale measurement on a 
dataset containing \finished{2.5M} sets of signals and \finished{1.4M} apps (the difference
between both figures is due to the same app
being distributed across different markets).
Second, we gather app and market
metadata (when available) from five Android app markets
between December 2019 and October 2021
(\S\ref{sec:dataset}). The resulting dataset constitutes the basis of our
study, whose main contributions are:

\begin{itemize}[leftmargin=*]
\item We empirically study the factors that impact on accurate
author attribution at the app and market levels, both within and across
Android markets	(\S\ref{sec:signals-consistency-on-markets}).
We measure and demonstrate that the use of market metadata and
signing certificates as attribution signals is unsound due to
the inaccuracy, volatility, and incompleteness of the data
(\rqoneref).

\item We introduce
the notion of an \emph{attribution graph} (\S\ref{sec:signal-consistency}) to study
signal consistency. By applying this concept to our datasets, 
we observe that attribution signals often conflict with
each other both at the app-level
(same author using different signals in different apps) and across market
(same app using different signals in different markets). These
conflict impede the accurate 
identification of app authors (\rqtworef).

\item We conduct a case study of the Google Play 
Store (\S\ref{sec:signals-consistency-on-the-play-store}). 
We demonstrate that its app vetting process is unable to detect 
forged metadata during the publication process despite Google's 
strict publication policies. Our work reveals that (1) the belief
that the signing certificate relates to a single company is invalid;
and (2) even a combination of several signals is insufficient for
sound attribution in the Google Play Store (\rqthreeref).
\end{itemize}

We conclude with a discussion of the 
scientific, operational, and regulatory implications of our findings.
We argue that the lack of supervision over Android
apps' release and signing process not only hinders both software 
transparency and accountability, but also impedes
developer-oriented measurement studies.
We believe that effective platform control is necessary to
solve the predicament of attribution on Android,
\eg{} by moving away from self-signed certificates and introducing
a trusted authority (\S\ref{sec:discussion}).

\vspace{1mm}
\noindent\textbf{Code and data.}
To foster reproducibility and further research, we provide both the source
code of our crawler, and the app and market metadata~\cite{datasetLink}.

     \section{Background}
\label{sec:background}

Both the software development
and the release processes implemented by every Android market involve multiple
parties. We note that the \textbf{owner} or
\textbf{author} entity which is accountable for the product can be different
than the entity (or entities) that took part in its development
---\ie{} the \textbf{developer}, which could be a software factory---; and the one
releasing it on a
market (\ie{} the \textbf{publisher}. Each of these stakeholders can
leave their own fingerprint in the software and market presence, which can
translate into incongruous signals that confuse the attribution process.
Therefore, we define author attribution as follows:

\begin{definition}[Attribution]
  Author attribution is the ability of a user to determine which company is behind
  a given app and is thus accountable for this product.
\end{definition}

The following example illustrates the challenges
behind Android app attribution. In the app ``Punk Music Radio,''
the app's certificate is signed by ``Andromo App'' (a development framework)
and the developer name
on the Play Store is a company called ``Yottabyte Enterprise Mobile.''
The package name of the app (\texttt{com.andromo.dev271569.app366038}) also points
to Andromo. However, the privacy
policy (\url{http://turtlefarmboost.simplesite.com/421262547}) is a broken website hosted on a domain
unrelated to any of these two companies. Given these mixed and contradicting signals, how can
we know which company is liable for this app? More generally, is it possible
to automatically and accurately
identify the authors of Android apps at market scale?
Even though there are legitimate use cases for introducing ambiguous attribution signals,
if this ambiguity is allowed without oversight, when it is not communicated to the end user
properly, or it hinders the ability for accurate 
attribution by analysts, it potentially becomes problematic.

The remainder of this section explains how
the development and signing process (\S\ref{sec:signing}),
and the publication of the Android
app through a market (\S\ref{subsec:market-policies-and-metadata})
can influence software attribution.

\subsection{Android App Signing Process}
\label{sec:signing}

Android apps are distributed as Android Package (APK) files through app
markets such as Google Play and alternative app stores like Huawei or Tencent. In order to provide \emph{integrity} (\ie{} to prevent
tampering of the content of the APK) and \emph{authenticity} (\ie{} to
prove the identity of the author), each package must be cryptographically
signed~\cite{google-play-recommended-certificate-lifetime}.
Google's
official policy states: ``\textit{the certificate associates the
APK [\dots] to you and your corresponding private key. This helps
Android ensure that any future updates to your app are authentic and come
from the original author}''~\cite{appSigningConsiderations}.
Therefore, the signing certificate supposedly plays a vital role as
an indication of authorship.

Android's signature scheme has been revised over
time.
For the sake of backward compatibility, APKs may be signed using one or more
signature schemes.
Most crucially, signing certificates typically are self-signed (99\% according
to one study on over 1M apps conducted in 2014~\cite{lindorfer2014andrubis}).
Yet, in contrast to Windows~\cite{kim2017certified} and
macOS/iOS~\cite{apple-mac-signing,apple-dev-enroll},
Android lacks a trusted authority that confirms the validity and
veracity of the certificates.

The Google Play Store also introduced new features with a direct
impact on the signing process and the ability to identify the
entity accountable for an app.
Since 2017, it offers
\textit{``Play App Signing''} as a way to protect signing keys from
being lost or
compromised by allowing app authors to delegate the key management and signing
process to the market. This service further constrains the already limited function of the signing
certificate as an indicator of authorship,
since these apps might be signed by Google itself and not by the
actual app author. In fact, since August 2021, Google requires all new apps to be published as app
bundles, which are essentially built and optimized by the Play Store and,
consequently, forces this signature delegation~\cite{bundles}.
In this case, authors can use self-signed certificates for
``\textit{inspection by developers and end users, who want to ensure that code
they're running matches the code that was originally built and signed by the
app developer}''~\cite{codeTransparency}.

Alternative Android markets
implement their own (typically less restrictive) policies.
For example, developers should upload an already-signed APK
to APKMonk~\cite{apkmonk-upload}, Baidu~\cite{baidu-upload} or
APKMirror~\cite{apkmirror-upload}.

\subsection{Market Metadata \& Policies}
\label{subsec:market-policies-and-metadata}

\begin{table}
  \centering
  \caption{Publishing policies for the 6 studied markets. Note that we could not find Baidu's policy.}
  \label{tab:policies_per_market}
\scalebox{.85}{\setlength{\tabcolsep}{0pt}
      \begin{tabular}{l ccccc}
                                    &
            \rot{35}{\textbf{\shortstack[c]{Google\\Play}}}   &
            \rot{35}{\textbf{APKMonk}}       &
            \rot{35}{\textbf{Tencent}}       &
            \rot{35}{\textbf{Baidu}}         &
            \rot{35}{\textbf{APKMirror}}     \\
          \midrule
          Publisher info & \y & \y & \y &  --- & \y   \\
          Anti-malware   & \y & \n & \n &  --- & \n \\
          Anti-clones    & \y & \n & \n &  --- & \y  \\
          False identity & \y & \n & \n &  --- & \y  \\
          \emph{Reference to policy}     &
            \cite{playStorePolicyCenter}  &
            \cite{apkmonk-policy}         &
            \cite{tencent-policy}         &
            ---                           &
            \cite{apkmirror-policy}       \\
    \end{tabular}
  }
\end{table}

Once signed, authors can publish their apps on one or several
markets. Most markets, with the exception of
Baidu,\footnote{Apps with the same package name can be published in
a game and non-game category.}
use the app package name for indexing apps.
Google's documentation stresses the importance of the package
name as an attribution signal and
recommends following the Java package naming convention (\ie{} to
``\textit{use Internet domain ownership as the basis for package names (in
reverse), to avoid conflicts with other developers}''~\cite{appPackage}).

App markets allow authors to disclose data about their app
(\eg{} its description and category) and themselves (\eg{} name and
contact information) in their market profiles.
However, publishing policies and profile
metadata are not consistent across markets
as shown in Table~\ref{tab:policies_per_market}.
Furthermore, the markets' terms of services (ToS) might set policies that can
influence author attribution. Specifically,
they often contain explicit policies to prohibit
impersonating other authors or distributing malicious
software and clones.
For example, Google's Developer Program
Policies~\cite{GooglePlayPolicyCenter}
prohibit misrepresentation and impersonation of other apps and
developers~\cite{playStoreMisrepresentation,playStoreImpersonation}.
These restrictions also apply to the
market metadata, such as the developer name,
title and screenshots.
However, it is unclear if
markets enforce these policies so the accuracy of this
data depends on the authors' will to honor the platforms' best practices
and guidelines.
Out of the markets that we study, only Tencent requires proof of identification
for registration, but only for Chinese citizens, residents, or companies.
We do not consider F-Droid in our study as it does not impose any requirements on authors. Because it downloads the source, F-Droid builds, and signs the apps itself so several apps exist with only a link to the code repository listed with no further metadata about the author.

     \section{Attribution Signals}
\label{sec:research_questions}

\renewcommand{\thefootnote}{\fnsymbol{footnote}}
\begin{table}[t!]
    \centering
\caption{Attribution signals available per market. The symbols
    denote the origin of the signal: the app's manifest (\frommanifest{});
    the app's signing certificate (\fromcertificate{}); market metadata (\frommarket{}).}
    \label{tab:signal_per_market}
\scalebox{.85}{\setlength{\tabcolsep}{2pt}
        \begin{tabular}{ll p{2.5em} p{2.5em}p{2em}p{2em}p{2em}p{2em}}
          & \rot{0}{\textbf{Origin}} &
            \rot{55}{\textbf{\# Markets}} &
            \rot{55}{\textbf{\shortstack[c]{Google\\Play}}} &
            \rot{55}{\textbf{\shortstack[c]{\quad\\APKMonk}}} &
            \rot{55}{\textbf{Tencent}} &
            \rot{55}{\textbf{Baidu}} &
            \rot{55}{\textbf{APKMirror}} \\
\midrule
            Package name       & \frommarketandmanifest{} & 6                             & \y                             & \y                         & \y                         &                          & \y                                                    \\
            App name           & \frommarketandmanifest{} & 6                             & \y                             & \y                         & \y                         & \y                       & \y                                                    \\
            Developer name     & \frommarketandcert{} & 5                             & \y                             & \y                         & \y                         &                          & \y                                                    \\
            Developer website  & \frommarket{} & 2                             & \y                             &                            &                            &                          &                                                       \\
            Privacy policy URL & \frommarket{} & 1                             & \y                             &                            &                            &                          &                                                       \\
            Developer email    & \frommarket{} & 2                             & \y                             &                            &                            &                          &                                                       \\
            Developer address  & \frommarket{} & 1                             & \y                             &                            &                            &                          &                                                       \\
        \end{tabular}
    }

\end{table}
\renewcommand{\thefootnote}{\arabic{footnote}}

A thorough analysis of (i) Google's official documentation; (ii) prior research 
in this area (see \ref{sec:sota}); 
and (iii) the app signing and release policies
across five Android markets, allowed us to identify four potential attribution
signals: (a) the package name, (b) the app name,
(c) developer details from market profiles; and (d) the signing certificate.
However, the availability of these signals across markets
is not consistent as we show in Table~\ref{tab:signal_per_market},
with the exception of
the signing certificate, which is mandatory for any Android app.

\parax{Package name}
The package name serves as the unique (string) identifier of an
app for the Android OS.
All markets use the package name as a unique and visible identifier,
or as an internal identifier, meaning that its value has to be unique
per app within each market, and should follow
Android's naming convention described in \S\ref{sec:background}.
One example of an app correctly following this naming convention is Facebook
(\texttt{com.facebook.katana}). However, since this convention is not enforced, many apps
introduce their own naming schemes as in the case of those
built with Andromo's development platform
(\eg \texttt{\{com,net\}.andromo.dev<dev\_id>.app<app\_id>}).

\parax{App name}
In some cases, the app name can be considered an
attribution signal itself (\eg Facebook). Howeer, it is not very robust
as this field might neither be
directly connected with its author nor be unique
(\eg consider generic names such as ``Music Player''). Despite its weakness, we
include this signal in our analysis to assess its validity
and consistency since it can be
extracted either from the apps' market profile or from their
manifest file. 

\parax{Developer details}
Authors publishing apps on Android app
markets are identified by a \signal{developer name}
and, depending on the market, they can also disclose
contact information such as an \signal{email address},
a \signal{website}, or their  \signal{physical address}.
Additionally, they can provide
a \signal{privacy policy URL} which must contain legal and
contact information of the author as required by current
legislation~\cite{gdpr,eudirective,ccpa}.
Therefore, the
value of market developer data for attribution purposes depends on
the accuracy of the data disclosed by the author.
Moreover, its validity
varies depending on the intended usage---specifically, on
whether it is used to attribute two apps \emph{published on the same market
  (\ie intra-market attribution)}, or
\emph{across markets (\ie inter-market attribution)} to the same author.

\parax{Signing certificate}
Since the private key associated with the certificate is
supposedly kept secret by its owner, there is a general belief
that two apps signed by the same certificate belong to the same
author. Each X.509 certificate contains a \signal{subject} field, which indicates the
\emph{owner} of the certificate, and an \signal{issuer}, which serves as
an indication of the entity that provided the certificate to the owner.
However, the subject
and issuer field in self-signed certificates are the same
and this information is filled in arbitrarily
by the party creating the certificate.
In addition, for apps that delegate their signing process
(see \S\ref{sec:signing}), all certificates generated by each platform share the
same subject field.

\subsection{Attribution Challenges in Prior Research}
\label{sec:sota}

Researchers have used the signing certificate, market
metadata or a combination of both for attribution.
However, no prior study has analyzed
the pitfalls of such attribution techniques and
how their accuracy
could compromise the interpretability and validity of results. Only the
limitations of signature-based attribution have been reported
in prior work.
An empirical analysis by Barrera \etal~\cite{Empiricalinstallation2012} on the
security aspects of Android app sandboxing
relied on certificates and app information as a proxy for authorship.
They showed that malicious actors often create
multiple self-signed certificates to prevent link-ability.
Similarly, Oltrogge \etal~\cite{oltrogge2018rise} showed that app building
frameworks, which automate app development and distribution, invalidate the
assumption that apps with the same certificate belong to the same company.

A particularly relevant line of research are market-level measurements and cross-market comparative
analysis~\cite{ali2017marketcompare, Zhong2013longtail,
viennot2014measurement, Holzer2011devperps, Heureuse2012whatsanapp,thomasimc2013}.
All of these studies relied on attribution signals such as the developer's
name to link app authors across markets.
Wang \etal{}~\cite{explorativemobile2017, wang2018beyond, wangwww2019} used
market-level metadata and certificate information to
track authors across Chinese app markets and the Google Play Store.
More recent efforts analyzed the Android supply chain and firmware-level
customizations, also faced attribution
challenges~\cite{gamba2020analysis,edu2021fota,vendorcustom2013}.
Specifically, Gamba~\etal reported that many preloaded apps are signed with Debug
certificates (thus violating Google policies)  or with vague subjects such
as ``Android,'' thus impeding attribution~\cite{gamba2020analysis}.

The security and privacy community has also used
app-level information for attribution purposes
~\cite{lindorfer2014andradar,findingmalice2015,groupdroid2017,rastogi2013appsplayground,
Ren2018bugfixes}.
A crucial aspect in security research is the ability to
responsibly disclose vulnerabilities. As most authors often
rely on market metadata (\ie developer address or privacy policies)
to contact app
developers~\cite{nguyen2021consent,incentivized2020farooqi},
the correctness of developer information is critical for accountability.
However, due to the inability to access accurate contact information
easily, researchers often opt to disclose their findings
to market operators like Google instead~\cite{li2021android, reardon201950}.
In other cases, researchers disclose a vulnerability in Android itself to
Google directly~\cite{tuncay2018resolving} and could use developer information
to inform affected app developers.

\begin{table*}[t!]
    \centering
    \caption{
        Dataset overview.
The overlap column reports the number of unique package names
        (PkgN) collected in both crawls.}
    \label{tab:count_overview}
    \scalebox{1}{\begin{tabular}{lrrrrrrrr}
            \multirow{2}{*}{\textbf{Market }} & \multicolumn{3}{c}{\textbf{First crawl (December 2019 -- May 2021)}} & \multicolumn{3}{c}{\textbf{Second crawl (June 2021 -- October 2021)}} & \multirow{2}{*}{
                \textbf{PkgN Overlap (\%)}} \\\cmidrule{2-7}
            & Market entries & PkgN    & APK SHA-256 & Market entries & PkgN    & APK SHA-256 &         \\\cmidrule{1-8}
            Google Play & 903,385        & 661,421 & 66,2019     & 540,361        & 434,884 & 440,060     & 150,731 ( \databar{18.746681}) \\
            APKMonk     & 324,054        & 298,133 & 300,341     & 500,870        & 293,351 & 293,469     & 153,472 ( \databar{26.481464}) \\
            Tencent     & 182,492        & 126,994 & 127,354     & 20,319         & 4,379   & 4,049       & 368     ( \databar{0.284920}) \\
            Baidu       & 5,235          & 4,304   & 3,881       & 13,907         & 2,533   & 2,536       & 38      ( \databar{0.481683}) \\
            APKMirror   & 752            & 716     & 722         & 2,846          & 1,113   & 1,113       & 6       ( \databar{0.329128}) \\
        \end{tabular}
    }
\end{table*}

     \section{Dataset}
\label{sec:dataset}

We use a custom-built Scrapy crawler~\cite{scrapy} to download APKs 
and their corresponding app metadata (when available) from five
markets (see Table~\ref{tab:signal_per_market}): 
Google Play Store, APKMonk~\cite{APKMonk},
APKMirror (an aggregator of apps published 
in other markets~\cite{APKMirror}),
and two markets with a large user base in
Asia: Baidu~\cite{baidustore} and Tencent~\cite{tencentstore}. 
We refer to the full set of signals extracted from each app listing 
in a market as a \emph{market entry}.

\parax{Crawling strategy}

We kickstarted each market crawl with a seed of all unique package
names from AndroZoo~\cite{androzoo2016}, totaling about 5M package names. Since AndroZoo does not link APK files
with the associated market metadata, we crawled the apps and metadata separately.
We opportunistically explore the six markets by following
links to similar or recommended apps shown in a given app's profile page.
Similar crawling strategies have proven effective
in prior work~\cite{wang2018beyond,ikram2016analysis}. We
deployed the crawler on two separate occasions (using 
the same 5M apps as a seed) between December 2019 and October 2021 to (1)
discover newly published apps, and (2) collect new version
of market entries for a subset of apps.
We extended our crawler between the two crawling windows to collect the
developer address from Google Play. 
We relied on twelve HTTP proxies located in two EU countries to parallelize the crawling.

\parax{Dataset statistics}
Table~\ref{tab:count_overview} provides an overview of our dataset. 
The first crawl represents a snapshot of apps collected between 
December 2019 and May
2021. The second crawl represents the snapshot of apps collected between June 2021
and October 2021. These two snapshots give us a longitudinal perspective of
attribution challenges in Android markets. 
Overall, we crawled metadata for \finished{1.36M} different package names across all markets and
\finished{1.45M} unique APK files (as identified by their SHA-256 file hash), covered by a total of \finished{2.49M} market entries.
We cannot evaluate the coverage of our crawler due to   
the inability to know the markets' full catalogue. 
The discrepancy
between the number of market entries and APKs crawled across markets
can be attributed to
artifacts such as rate limiting, timeouts, geo-restrictions, and other errors
when downloading the APKs. Even though we used multiple IP addresses for the
crawler, we were unable to crawl
APKMirror at scale, as it lies behind CloudFlare's DDoS protection system.
Still, for Google Play, we collected \finished{1.40M} market entries with
\finished{804k} unique apps out of the estimated \finished{2.65M} available at the time of
writing~\cite{gplay-stats}.

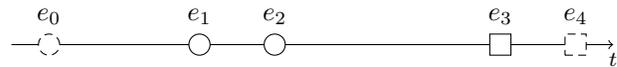
\begin{figure}[t]
    \centering
    \begin{tikzpicture}
        \tikzstyle{circ}=[circle, draw, fill=white, minimum size=8pt, inner sep=2pt]
        \tikzstyle{sqr}=[rectangle, draw, fill=white, minimum size=8pt, inner sep=2pt]
        \draw[->] (-0.5, 0) -- (7.5, 0);
        \node at (7.5,-.2) {\footnotesize $t$};
        \node[circ, label={$e_0$}, densely dashed] at (0,0) {};
        \node[circ, label={$e_1$}] at (2,0) {};
        \node[circ, label={$e_2$}] at (3,0) {};
        \node[sqr, label={$e_3$}] at (6,0) {};
        \node[sqr, label={$e_4$}, densely dashed] at (7,0) {};
    \end{tikzpicture}
    \caption{Timeline of the collected market entries for a given package name and market
        ($\circ$ = first crawl, $\square$ = second crawl).
        The longitudinal analysis investigates the differences between $e_0$ and $e_4$,
        whereas the other analyses consider $e_4$ only.
    }
    \label{fig:market_entry_selection}
\end{figure}

\parax{Longitudinal analysis}
Performing a longitudinal analysis requires us to handle both different versions
of apps and changes in their market profiles. 
Table~\ref{tab:count_overview}
reports the overlap of package names across markets 
for our two crawling campaigns. The
overlap varies from market to market due to app removals 
and our best-effort crawling strategy: we note
that some packages were unlisted from the stores between crawls. Still, the
second crawls for APKMonk and Google Play contain \finished{19\%} and \finished{27\%} of the apps
collected in the first one, respectively.
While apps and market metadata are
frequently updated, these changes might occur at different times,
\ie not every market listing for a given package
name results in a different APK file.
This phenomenon explains the mismatch between app package names and APKs in
Table~\ref{tab:count_overview}.
In order to conduct our longitudinal 
analysis in such a variable
scenario (\S\ref{sec:signals-consistency-on-markets} and
\S\ref{sec:signals-consistency-on-the-play-store}), 
we use the latest collected market
entry except for our longitudinal
analysis (\S\ref{subsec:attribution-signal-volatility}), in which we consider 
the earliest entry from the first crawl and the latest from the second for each app
(see Figure~\ref{fig:market_entry_selection}).

     \section{Analysis of Attribution Signals}
\label{sec:signals-consistency-on-markets}

Each Android app market defines its own policies
and enforcement mechanisms 
for publishing apps (\S\ref{sec:background}). Yet, all attribution
signals are self-declared by the author or developer of the
software, thus they could be missing or misleading.
Moreover, attribution data is not necessarily 
persistent. The Android ecosystem is highly dynamic 
with frequent company acquisitions, 
re-brandings, and new apps and versions
being released regularly~\cite{Ren2018bugfixes, calciati2018did}, 
as well as short-lived impersonifications 
(App-Squatting)~\cite{Hu2020appsquating}.
As a result, attribution data might not be consistently updated, 
thus leading to confusion. In this section, we measure the \emph{availability and volatility} 
of attribution signals across markets 
in order to reason about their validity and coverage, answering \rqoneref.

\subsection{Signal Availability}
\label{subsec:missing-signals}
As a first step, we empirically measure to what extent individual 
attribution signals are missing on apps across
markets. 
This preliminary analysis is critical to assess the enforcement of
market-specific publication policies and then to reason about their individual validity.
Later in \S\ref{subsec:signal-consistency-market-app}\---\S\ref
{sec:signals-consistency-on-the-play-store}, we analyze combinations of
signal values and how the soundness of such approaches is negatively impacted
when markets fail to provide signal values for published apps.

For this analysis, we consider  
the latest version for each app (per market listing) in our dataset.
Table~\ref{tab:missign_signals} shows the percentage of unique package
names for which there are missing signals, 
across all markets in our entire dataset.
For market entries from Google Play, several signals were not
collected for the full data collection period and we report  (\S\ref{sec:dataset}) those
results as a percentage of packages collected after we started collecting them.

\parax{Market metadata}
Nearly all market entries collected across the markets
are published under a developer name.
Of the eight app entries 
on Google Play without a developer name, 
only one (\texttt{br.com.mksolutions.mksac.redebr})
remains listed at the time of writing.
More significant, however, is the extent to which developer
websites (\finished{34}\%), developer addresses (\finished{44}\%), and privacy
policy URLs (\finished{18}\%) are
missing from Google Play profiles, and the extent of signal absence on
the rest of the markets.
The contact information is clearly not seen as a necessity for app publication on the Play Store,
neither by Google nor the developers.

\parax{Signing certificates}
When looking at the signing certificates' subject, we find that relative
distinguished name (RDN) components are missing for a significant fraction of
unique market entries,
ranging from \finished{7}\% for the common names on Google Play to
\finished{39}\% of the state missing on APKMonk.
In Android, the RDNs do not serve any purpose for app developers, hence
omitting them from their signing certificates comes with no drawbacks.
The lack of oversight on the self-signed certificates results in the
inconsistent availability of these RDNs.

\renewcommand{\thefootnote}{\fnsymbol{footnote}}
\begin{table}[t]
    \centering
    \caption{Percentage of unique market entries throughout the dataset with missing attribution signals on the
    different markets (\notcollectedsignal{} indicates we did not collect a specific signal).}
\label{tab:missign_signals}
    \scalebox{.95}{\setlength{\tabcolsep}{2pt}
        \begin{tabular}{llrrrrrr}
            \multicolumn{2}{c}{\textbf{Attribution Signal}} &
            \multicolumn{1}{c}{\rot{55}{\textbf{\shortstack[c]{Google\\Play}\footnotemark[1]}}} &
            \multicolumn{1}{c}{\rot{55}{\textbf{APKMonk}}}       &
            \multicolumn{1}{c}{\rot{55}{\textbf{Tencent}}}       &
            \multicolumn{1}{c}{\rot{55}{\textbf{Baidu}}}         &
            \multicolumn{1}{c}{\rot{55}{\textbf{APKMirror}}}     \\
\midrule
            \multirow{5}{*}{\rot{90}{\textbf{Market}}}
            & Developer name              &         <0.01\% &                <0.1\% &               <0.01\% & \notcollectedsignal{} &                   0\% \\ & Developer website           & \databar{33.71} & \notcollectedsignal{} & \notcollectedsignal{} & \notcollectedsignal{} & \notcollectedsignal{} \\ & Developer email             &         <0.01\% & \notcollectedsignal{} & \notcollectedsignal{} & \notcollectedsignal{} & \notcollectedsignal{} \\ & Developer address           & \databar{44.43} & \notcollectedsignal{} & \notcollectedsignal{} & \notcollectedsignal{} & \notcollectedsignal{} \\ & Privacy policy URL     & \databar{17.75} & \notcollectedsignal{} & \notcollectedsignal{}      & \notcollectedsignal{} & \notcollectedsignal{} \\ \midrule
            \multirow{6}{*}{\rot{90}{\textbf{Cert RDN}}}
            & \texttt{commonName}         &  \databar{7.12} &        \databar{8.97} &        \databar{9.08} &  \databar{6.54} &       \databar{11.90}  \\ & \texttt{organization}       & \databar{17.60} &       \databar{25.38} &       \databar{21.11} &  \databar{20.93} &        \databar{8.45} \\ & \texttt{org.Unit}           & \databar{27.64} &       \databar{36.69} &       \databar{30.87} &  \databar{21.92} &       \databar{23.53} \\ & \texttt{locality}           & \databar{27.22} &       \databar{35.85} &       \databar{29.27} &  \databar{20.92} &       \databar{16.90} \\ & \texttt{state}              & \databar{28.78} &       \databar{38.79} &       \databar{32.64} &  \databar{24.27} &       \databar{20.30} \\ & \texttt{country}            & \databar{21.79} &       \databar{30.34} &       \databar{27.00} &  \databar{24.01} &       \databar{15.41} \\ \end{tabular}
    }
	\caption*{\footnotemark[1]{\scriptsize Developer name, address and email as \% of markets entries crawled after 2021/09/21}} 

\end{table}
\renewcommand{\thefootnote}{\arabic{footnote}}
 
\subsection{Signal Volatility}
\label{subsec:attribution-signal-volatility}
We measure volatility of market metadata and signing certificates by
monitoring changes across the two dataset snapshots as
described in \S\ref{sec:dataset}:
we select the earliest entry from the first snapshot and its
latest entry from the second snapshot (see Figure~\ref{fig:market_entry_selection}).
Overall, we observe that
the coverage of package names varies widely across markets, ranging
from \finished{less than 1\%} for Tencent, Baidu and APKMirror to \finished{27}\% for APKMonk.
We report the results for the three low-coverage markets, but note that the results are not representative for the market as a whole.
Table~\ref{tab:signal_volatility} shows
the results for the signals collected in both crawls.

\parax{Market metadata}
All markets allow app and developer names of published apps to change: the
app name volatility ranges from \finished{0.9}\% of package names on both
\finished{APKMonk} to \finished{7.9}\% of package names on \finished{Baidu}.
The volatility of the developer name ranges from \finished{0.6}\% on
\finished{APKMonk} to \finished{18.8}\% on \finished{Tencent}.
For the signals captured on Google Play only (\ie the
developer website, email address, and privacy policy URL) we see a
volatility of \finished{1.7}\%, \finished{1.0}\%, and
\finished{1.6}\%, respectively.

Notable cases of highly volatile signals include the app `Electronic Dance Music Radio' and the developer `Cube Apps Ltd' who underwent
six and four name changes over the course of our measurement setup respectively.
Note that these results do not include cases in which signal values differed 
during the measurement period
but ended up with their original values.
A manual inspection suggests that such cases occur only sporadically 
and nearly all of these identified cases
consist of a single change that was reverted.

Changing the information under which an app is published has its legitimate use cases,
but our results suggest that volatility should be considered as
a potential issue in research studies.
Conclusions drawn from signals
at a certain point in time may not hold at a later stage for a given body of
package names.

\parax{Signing certificates}
Signing certificates are volatile too: \finished{58}\% of all apps
in our dataset are signed with multiple certificate schemes (such as v1, v2, v3).
Specifically, \finished{23.1}\% of the apps are signed with v3
certificates, hence allowing certificate rotation (the number rises to \finished{29.1}\% on Google Play).
The certificate percentages in
Table~\ref{tab:signal_volatility} represent the fraction of package names that
added a new certificate to the APK or completely changed
certificates across their market entries. The latter can contain re-published apps
by either their original or by a completely different author.
The self-signed nature of the signature does not allow us to
differentiate between these two
cases due to the lack of a ground truth.
For example, our dataset contains two market entries
for package name ``\texttt{ua.iread.android}''
on Google Play, published under a different developer name
(``\emph{<pineconeapps>}'' and ``\emph{xl-games}'') and different app names
(``\emph{Learn to read for kids free}'' and its seemingly Ukranian translation).
The different developer names do not provide any confidence that both
apps were published by the same organization.

We also find apps from different authors that have a signature that
suggests the delagation of the signing process to Google through ``Play App Signing``.
However, the lack of ground truth prevents us
from certifying whether the app was signed by the Play Store or the publisher
just created a signature with the same subject information as the Play Store's.
We find that this happens in all markets, ranging from \finished{less than 1\%}
of package files on \finished{Baidu} to \finished{16.0}\% on APKMonk and \finished{32.4\%} on \finished{Google Play} itself.
These exclude apps signed by certificates with a similar subject to the
Google Play's default subject field (\eg{} the developer 'Yippity Doo LLC'
publishing an app using a certificate with ``\emph{Google}'' and
``\emph{Cupid}'' as the subject organization and common name,
instead of ``\emph{Google Inc.}'' as the common name used by the Play Store
for their certificates), which are apps published under bogus certificates.
Without analyzing additional signals extracted from the APK, this ambiguity
affects the soundness of studies into signing delegation.

\begin{table}[t]
    \centering
    \caption{Percentage of package names per market for
    which we observe a change in signals over time. App coverage reports
    the percentage of apps present in both crawls.}
\label{tab:signal_volatility}
    \scalebox{.95}{\setlength{\tabcolsep}{2pt}
        \begin{tabular}{llrrrrrrr}
            \multicolumn{2}{c}{\textbf{Attribution Signal}} &
            \multicolumn{1}{c}{\rot{55}{\textbf{\shortstack[c]{Google \\Play}}}} &
            \multicolumn{1}{c}{\rot{55}{\textbf{APKMonk}}} &
            \multicolumn{1}{c}{\rot{55}{\textbf{Tencent}}} &
            \multicolumn{1}{c}{\rot{55}{\textbf{Baidu}}} &
            \multicolumn{1}{c}{\rot{55}{\textbf{APKMirror}}} \\ \midrule
            \multirow{5}{*}{\rot{90}{\textbf{Market}}}

            & App name           & \databarfull{2.5} & \databarfull{0.9}     & \databarfull{6.8}     & \databarfull{7.9} & 0\% \\
            & Developer name     & \databarfull{3.9} & \databarfull{0.6}     & \databarfull{18.8}    & \notcollectedsignal{} & \databarfull{16.7} \\
            & Developer website  & \databarfull{1.7} & \notcollectedsignal{} & \notcollectedsignal{} & \notcollectedsignal{} & \notcollectedsignal{} \\
            & Developer email    & \databarfull{1.0} & \notcollectedsignal{} & \notcollectedsignal{} & \notcollectedsignal{} & \notcollectedsignal{} \\
            & Privacy policy URL & \databarfull{1.6} & \notcollectedsignal{} & \notcollectedsignal{} & \notcollectedsignal{} & \notcollectedsignal{} \\

            \midrule
            \multirow{2}{*}{\rot{90}{\textbf{Cert}}}
            & Added              & 0\%               & 0\%                   & \databarfull{0.3}     & 0\%                   & \databarfull{16.7}    \\
            & Fully replaced     & <0.01\%           & <0.01\%               & \databarfull{2.7}     & 0\%                   & 0\%                   \\

\end{tabular}
    }
\end{table}

\vspace{1em}
\noindent\textit{\textbf{RQ1: \rqone}}
Our empirical results highlight that the developer contact 
signals are relatively unreliable for attribution,
either because they are not available at all or, in the case of Google Play, often missing.
The developer name \--- besides the app name and package name \--- is essentially 
the one reliably available signal in app markets.
The lack of imposed and enforced restrictions on the (self-signed) certificates is visible in the
fraction of RDNs that are missing across signing certificates.
Furthermore, whereas certificate remain relatively stable over time, 
market signals tend to be more volatile.

     \begin{figure}
    \centering
    \includegraphics[width=.9\linewidth]{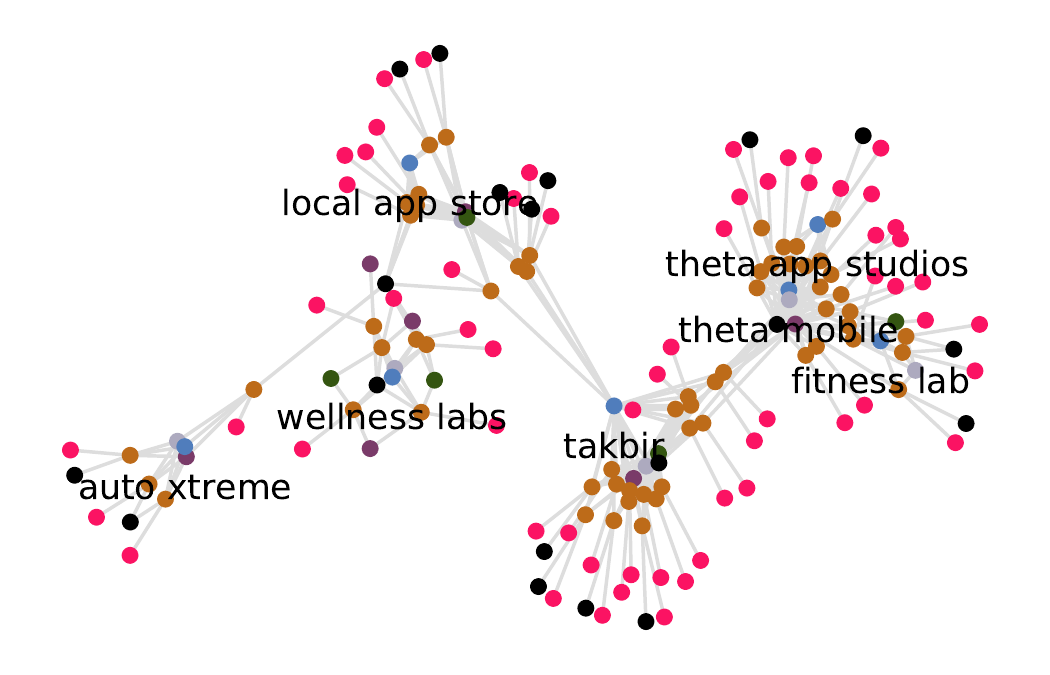}
    \caption{A cluster from the Google Play attribution graph annotated with developer names (\tikzcircle[marketentrycolor,fill=marketentrycolor]{0.2em}~=~market listing,
        \tikzcircle[developernamecolor,fill=developernamecolor]{0.2em}~=~developer name,
        \tikzcircle[developerwebsitecolor,fill=developerwebsitecolor]{0.2em}~=~developer website,
        \tikzcircle[developeremailcolor,fill=developeremailcolor]{0.2em}~=~developer email,
        \tikzcircle[privacypolicycolor,fill=privacypolicycolor]{0.2em}~=~privacy policy URL,
        \tikzcircle[appnamecolor,fill=appnamecolor]{0.2em}~=~app name,
\tikzcircle[certcolor,fill=certcolor]{0.2em}~=~certificate).}
    \label{fig:attribution_graph_messy}
\end{figure}

\section{Signal Consistency}
\label{sec:signal-consistency}

Beside the availability and volatility of individual signals, their
self-declared nature and lack of market enforcement directly affects their
consistency.
The signals associated with an app published by a given
developer in a given market (and across markets)
may conflict with one another,
thus potentially impacting attribution efforts.
By evaluating these conflicts, we answer \rqtworef~in this section.

\subsection{Attribution graph}
We introduce the novel concept of \textit{attribution graph}
to study and measure signal consistency at the app level
(\S\ref{subsec:signal-consistency-market-app}), within markets
(\S\ref{subsec:signal-consistency-within-markets}), and
across markets (\S\ref{subsec:signal-consistency-across-markets}).

\begin{definition}[Attribution Graph]
  We define an attribution graph as a bipartite graph $G=(S, A, E)$ in which:
  \begin{itemize}
      \item $S$ is the set of market entries, \ie the set of all the different signals that can be extracted for
      each market listing (as described in \S~\ref{sec:research_questions}).
      These are uniquely identified by the lowercase value of
      most signals (\ie developer name, website and email, the privacy policy URL, package name
      and the app name) and by the SHA256 digest of the signing certificates.
    \item $A$ is the set of all market listings, \ie every unique
      collection of metadata for a given app and market.
    \item $E = \{ (x,y) \mid x \in S \land y \in A \}$ is the set of edges
      that connect signals with the market entries in which they are found.
   \end{itemize}
\end{definition}

This attribution graph allows extracting and measuring
the relationship between the different signals and market listings, and their
consistency by extension, by analyzing different
subgraphs. 
Figure~\ref{fig:attribution_graph_messy} illustrates this concept and
its application to study signal consistency
by showing a cluster of apps published on Google Play
apps. 

\subsection{Signal Consistency Within Apps}
\label{subsec:signal-consistency-market-app}
Due to the fact that attribution signals can be potentially introduced by different actors
at different stages of the publication process  (\S\ref{sec:research_questions}),
it is possible that this information is not consistently introduced or updated
for a given app. 
One such examples is the app name, which is present both in the market and in 
the APK file. 
We measure the prevalence of such inconsistencies 
by comparing the app name disclosed in the market metadata with the
information present in the manifest file 
for every market listing of a given app (per package name). 
We represent the market listing and its two app names in our attribution graph
as nodes, and compare the label of the app name nodes.

App authors might use other alphabets, leading to potential mismatches if
not handled correctly. 
Therefore, to avoid bloating our results,
we discard pairs of market and app data where one of the two signals 
is not in the
Latin alphabet. For the remainder of apps, we observe that only in \finished{45.3}\%
of the cases the name is exactly
the same. All markets suffer from these inconsistencies \--- ranging from
\finished{28.9\%} and \finished{46.0\%} on APKMirror and Tencent respectively \---
which hints towards poor software maintenance practises and lack of enforcement by
market operators.

Such signal inconsistencies complicate the comparison
of research studies and attribution
efforts at the app-name level. In fact, they make
external analysis of these apps harder as one has to decide 
whether to rely on market metadata or the app's manifest 
(\eg when researchers make their results
available to the public). Furthermore, the possibility to 
create an inconsistency between the market and the 
installed app can be abused by malware and phishing
attacks~\cite{malware:titlecionchange}. Users might be tricked into downloading
a potentially privacy-intrusive app that shares the name with a
well-known app~\cite{clubhouse, removed2018wang}.

For the apps for which we do
not find an exact match, we measure how similar both names are. To do so, we rely on the
normalized Levenshtein similarity~\cite{marzal1993normalizededitdistance}. 
We find that around \finished{17.4}\% of the
apps have a similarity below \finished{50}\%. We observe cases in which one name
is a substring of the other (\eg ``\emph{racing lap timer \&
stopwatch}''---``\emph{laptimer}'',
``\emph{localwifinlpbackend}''---``\emph{wifi location service}''), but also cases
in which both names appear to be completely
unrelated (\eg ``\emph{marshmallow adventure}''---``\emph{flappy candy}'' or
``\emph{filebox}''---``\emph{myfaves}'').
The package name and app icon of ``\emph{marshmallow adventure}'' reveals
that the app was previously promoted under the name in the app manifest as a
clone of the well-known \emph{Flappy Bird} app,
whereas the origin of ``\emph{myfaves}'' is unclear.
Our findings confirm the belief that relying on its name to identify an app
(as users tend to do~\cite{clubhouse}) can lead to the
installation of undesired apps. We also show the potential applications of
attribution graphs to detect clone apps.
 
\subsection{Signal Consistency Within Markets}
\label{subsec:signal-consistency-within-markets}
We now assess signal consistency
within a given market from an author perspective, \ie{} the extent to which the
same signals are used across all apps by a publisher. 
The company developing
an app can be different to the app's author and even
to the company publishing it (see \S\ref{sec:background}).
Still, signal consistency is critical for researchers, users, and 
regulators to correctly and unambiguously attribute authorship.

We evaluate the consistency by measuring the re-use of signals by a particular
publisher, as well as across publishers, focusing on those
signals available on most markets: the app name, the developer name
(per market metadata), and the signing certificate.
Since these signals are not all available on Baidu, we exclude it from the analysis.

First, we evaluate the degree to which certificates are uniquely 
used by a single publisher.
We analyze our attribution graph, identifying all developer name
nodes that are connected to a particular certificate fingerprint node,
which represents the number of developer names publishing apps signed
with this certificate.
We find that a relatively small number of certificates on the markets are used
across multiple developer names (ranging from \finished{1.2\%} on
\finished{Google Play} to \finished{4.9\%} on \finished{Tencent}).
However, these certificates are used to sign a large fraction of market entries
(ranging from \finished{15.2\%} on \finished{Google Play} to \finished{22.6\%} on
\finished{Tencent}).
Some of these market entries are associated to apps from the same
developer that are published under different developer names 
(\eg international subsidiaries, business units, or development teams),
but most cases
are associated with app building frameworks like Andromo~\cite{andromo} or
AppyPie~\cite{appypie}.
These development frameworks offer software authors
mechanisms for outsourcing the app building process either through
automatic app creation techniques, 
or by developing the app for the publisher~\cite{oltrogge2018rise}.
The final app, however, is signed directly by the app building framework.
These artifacts result in many
unrelated apps sharing the same signing certificate. We study app building
frameworks further in \S\ref{subsec:multiple-signal-attribution-graphs}.

We also quantify the opposite aspect: the number of certificate
fingerprint nodes connected to a developer name which measures the number of
certificates used by the developer to sign their apps.
We find that it is common practice for authors to release apps with more
than one certificate: the percentage of developer accounts of apps signed with
multiple certificates ranges from \finished{9.3\%} for \finished{APKMirror}
to \finished{44.5\%} for \finished{Google Play}, and affect
\finished{34.2\%} and \finished{66.1\%} of market entries
on the two markets, respectively.
If two apps by the same developer operate fully independent, there is
currently no incentive for a developer to publish them under the same certificate.

Using the same method, we also group market entries by app name and
count the number of apps published under more than one developer name.
We find that the percentage of listings that share their app name 
with another listing but are published by different developers
varies from \finished{0\%}
on \finished{APKMirror} to \finished{3.4\%} on \finished{APKMonk}.
A notable example on Google Play is the app name \emph{Messages}, 
which has been published under \finished{25} different package names, 
with Google's version being the most popular one with more than 
1B downloads.

\begin{figure}
    \centering
    \includegraphics[width=\linewidth]{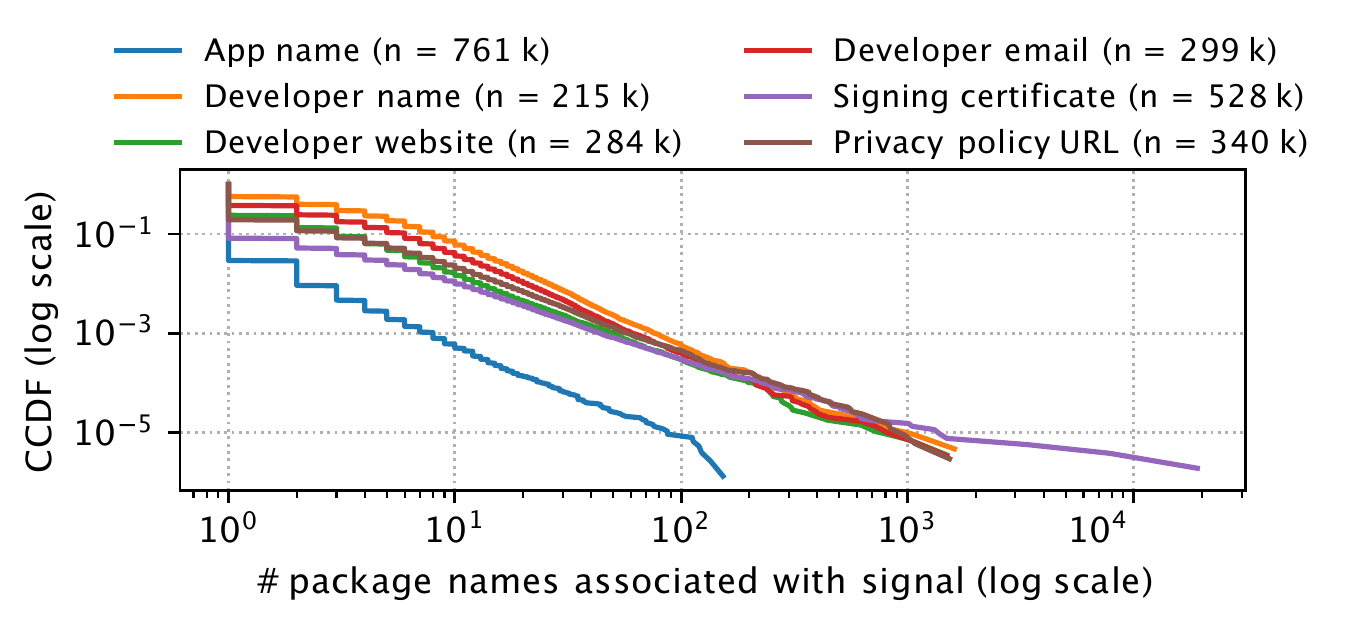}
    \caption{Complementary CDF of the number of package names associated with
    individual signals on Google Play.}
    \label{fig:ccdf_cluster_sizes_single}
\end{figure}

\parax{Google Play}
We direct our efforts towards analyzing those attribution 
signals that are only consistently available on Google Play:
the developer website, the privacy policy URLs, and developer email.
There is a relatively low correlation between the
developer name and other developer-related signals: 
\finished{26.3\%} of developer names publish apps
with more than one website, \finished{23.7\%} with multiple
email addresses, and \finished{31.2\%} with multiple privacy policies.
The latter is an expected outcome, as two apps by the same developer 
can collect different types of personal data and hence may 
have a different privacy policy. Note that the automatic analysis of the
privacy policies to extract attribution information is an orthogonal research
challenge~\cite{zimmeck2016automated,harkous2018polisis,andow2019policylint}. 
The extent to which email addresses and websites differ is more surprising.
The developer ``\emph{Rad3 Limited}'' published two apps ``\emph{Wakachangi GO}''
and ``\emph{Smales Farm}'' under the email and websites ``\emph{contact@wakachingi.com}''
and ``\emph{http://www.wakachangi.com/}'', and ``\emph{mark@smalesfarm.co.nz}''
and ``\emph{https://smalesfarm.co.nz}'').
These findings show that contact information is in some cases
\emph{app}-related and not \emph{developer}-related.

We further examine the distribution of package names published under a
particular signal value shown in Figure~\ref{fig:ccdf_cluster_sizes_single}
as a complementary cumulative distribution function (CCDF).
The figure shows that for each signal type, there are outliers of signals
under which a large volume of apps have been published.
These cases \--- which include the aforementioned \emph{Messages} app name \---
highlight the inability of relying on a single signal for sound attribution.

\subsection{Signal Consistency Across Markets}
\label{subsec:signal-consistency-across-markets}

An app's package name uniquely identifies an app within the Android OS
and within markets.
However, the association of the unique package name with a particular app might
not prevail from market to market
since a market operator can only enforce this policy on their own
market.

Signal inconsistency across markets
has implications in research efforts trying to compare the
catalogue of markets and developer publishing and release strategies:
the package name can be unreliable for identifying
unique apps across markets.
Hypothetically, two different entities sharing the 
same package name might publish an app each
under the same developer name but on two different markets.
The signals would suggest the same underlying developer even though this is
not the case, thus leading the user to install a different app (with a different
behavior) depending on the market where it is downloaded.
This is also important for researchers, as the market of origin becomes a
relevant variable when mining repositories and reasoning
about the data and measurement results,
or when studying market catalogues and their risks
for the end users. In other words, measurements on Google Play might not
extrapolate to other markets, or
cross-market datasets might be polluted by unrelated apps.

\begin{figure}
    \centering
    \includegraphics[width=.6\linewidth]{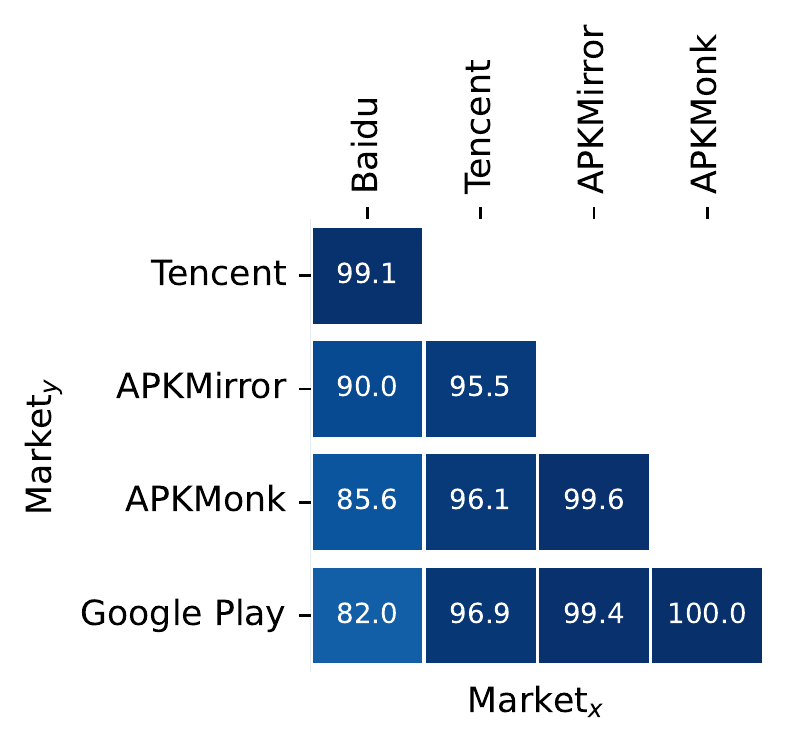}
    \caption{Percentage of package names published on both
    market\textsubscript{x} and market\textsubscript{y} signed with the same certificate.}
    \label{fig:signing_certs_across_markets_rel}
\end{figure}

We calculate the number of apps
(by package name) listed on more
than one market to measure how inconsistent attribution
signals are across markets. Out of the total \finished{1,355,186} unique
package names present in our dataset,
\finished{158,735} (\finished{11.71}\%) have been published on multiple markets.
We compare the certificate nodes in the attribution graph used to sign apps
across markets sharing a package name.
Figure~\ref{fig:signing_certs_across_markets_rel} shows the percentage
of these apps per market pairs that are signed by
the same certificate on both markets.
Notably, APKMonk and Google Play
have a full overlap.
This suggests that APKMonk operates as a mirror of Google Play.
The overlap between Baidu and the western markets is relatively poor (between 82\% and 90\%), which
raises the question whether market entries that share a package name are the same app and
are published by the same author.

For apps co-published on both Google Play and another market,
we see that between the percentage of apps signed by a Google certificate (as identified by its subject)
on the alternative market ranges between \finished{2.4}\% and \finished{24.3}\% for Baidu and Tencent
respectively.
Those are apps that are likely to have first been published on
the Play Store, after which their signed APKs were published on the other
market as recommended by Google~\cite{google-signing-key-upgrade}.
As such, Google Play's signing policy does not only affect their own market,
but has also started to impact on
attribution of apps across markets at the certificate level.

In a similar fashion as the certificates, we compute the
percentage of package names published under the same
app name between market pairs (\ie every possible combination of two markets).
We run this
analysis only for those app names sharing the same alphabet
to avoid noise caused by language differences.
We observe that there are significant differences in the app
names depending on the origin of the market.
We find that Chinese markets depict a relatively higher overlap of app names.
\finished{77.0}\% of package names published on a pair of Chinese markets are
published under the same app name. The percentage is even higher for pairs of
Western markets (\finished{91.7}\%). However, 
the percentage is much lower for those package names published
across a Western and a Chinese market
(\finished{52.7}\%). 

Similarly, we compute these percentages for the developer
name. We exclude Baidu from this analysis since this market does not report developer
names (\S\ref{sec:dataset})).
Across Western markets, package names are published under
the same developer name in
\finished{92.2}\% of the cases, and only \finished{66.5}\% across Tencent and
Western markets. We find
examples of apps in which the developer name is related across markets (\eg
``\emph{naver webtoon corp.}'' and ``\emph{webtoon entertainment}''), while
others are completely unrelated (\eg ``\emph{gomo limited}'' and ``\emph{go
launcher dev team}'').
Possible explanations for the disparity between Chinese and Western signal values include
more rampant app repackaging, publication of apps under a subsidiary (or different)
company or poor practises of synchronizing metadata on markets across regions.
However, these hypothesis 
are difficult to validate with our dataset.

\vspace{1em}
\noindent\textit{\textbf{RQ2: \rqtwo}}
We identified a significant lack of overlap between app names declared in the
app's manifest and the published app name.
We find both a significant duplicate signal usage between
developer accounts (\eg identical certificates and
app names used to publish app from different developer accounts) and multiple
signal usage by a single developer account within markets
(\eg publishing apps under different certificates or email addresses).
Moreover, our results highlight the challenge of tracking authorship across
markets, due to their package names
being published under different app names and \-- perhaps more
importantly \-- different certificates. This problem is
magnified by tracking package names across Western and Chinese markets.

     \section{Case Study: Google Play Store}
\label{sec:signals-consistency-on-the-play-store}

Our results in \S\ref{sec:signals-consistency-on-markets}
and \S\ref{sec:signal-consistency} confirm 
that publicly available attribution signals are unsound and can lead
to confusion. 
Other signals available in market profiles, 
such as the developer website or their email address could improve attribution,
but these signals are only reliably available on Google Play.
The absence of these signals in most alternative markets 
not only harms attribution of app authors
(\ie transparency), but also automatic and large-scale accountability and comparative
market analysis. For example, users might not be able to 
easily find the author's contact information 
about the company prior to installation in order to decide whether
to trust or not the developer, or even 
exercise their data rights in case they need to. 
Similarly, researchers might not be able to 
responsibly disclose vulnerabilities without contact details.

In this section, we set out to answer \rqthreeref.
We focus our analysis explicitly on the Google Play Store,
as (1) it is the official
and most prevalent Android marketplace~\cite{gplay_prominence}; (2) it
is the richest in attribution signals at the market level; 
and (3) Google Play has a significant number of publication policies
(\S\ref{subsec:market-policies-and-metadata}) that implicitly or explicitly
affect attribution.

\subsection{Enforcement of Publication Policies}
\label{sec:vetting-process-of-the-play-store}

We start by investigating the enforcement of Google Play's
publication policies through different experiments involving
the publication of potentially deceptive apps on Google
Play. Particularly, we focus on those policies and recommendations
related to attribution signals. 
According to Google's documentation, 
each app submitted to Google Play goes through an 
automated vetting process that
flags an app for manual inspection if it is found to contain suspicious
behaviors~\cite{play-store-cloud-protection}. As impersonation and tampering
are explicitly prohibited (\S\ref{subsec:market-policies-and-metadata}),
apps violating these policies can be expected to be rejected by the
market. The primary goal of our experiment is to confirm this 
hypothesis and assess the robustness of
Google Play's vetting process to detect and prevent the publication of 
apps that attempt to impersonate
or tamper with other apps or developer profiles. In doing so, we also shed
light onto the fidelity of the attribution signals available in the
market.

For each experiment, we compiled and attempted to publish a repackaged
and completely benign app that includes partially fake developer 
information. We chose two open source apps with
licenses that allow their redistribution to avoid infringing on
any intellectual property rights. Namely, we leverage \emph{Open
  Sudoku}~\cite{sudoku},
a game with a moderate number of downloads (10,000+),
and the \emph{Signal Private Messenger}~\cite{signal},
a popular privacy-focused
messenger, with more than 100 million downloads as of March 2022.
Specifically, we tested whether Google Play (1) conducts
any code similarity analysis between new and existing apps;
(2) evaluates the similarities between market signals; and (3) carries
out a more exhaustive vetting if the affected app has a high profile
(\ie a larger number of downloads).
We released each app under a new developer account
to prevent apps from being rejected based on developer reputation.

\parax{Ethical Considerations}
We took several steps
to ensure that our experiments would cause no harm to users
or to the platform itself. We received approval from our
institutional ethical review board before conducting the
experiments. We note that our intention is not measuring whether users 
install clone apps but whether these apps get flagged during the
review process.  All apps that we published for this experiment
display a notification to the users, clearly stating
that they are part of a research project and they are not the original
one. Additionally, this notification redirected users to the 
original apps and
advised them to uninstall our clone. Furthermore, we did not collect
any personal information about the users that installed our apps.
Finally, we removed the apps from the Play Store after our 
experiments to prevent any confusion.
Our experiment might have affected the platform,
though we consider the impact to be negligible. Since all apps go through
an automatic vetting process when published, and with an estimated
100k apps~\cite{monthly-gplay} published monthly, the overhead
of two apps is minimal. There
are over 3 million apps~\cite{total-gplay} publicly 
available on the Google Play
Store, so the release of three apps (\ie one app was
released on two different occasions)
will not affect any potential measurement
or research conducted on the whole spectrum of
the Google Play Store in any meaningful way. Thus, 
we estimate the benefits of understanding the vetting process 
of the Google Play Store and its shortcomings to be far greater
than the potential overhead of publicly releasing two apps.

\parax{Experiment \#1: Code similarity}
Performing any code similarity for every new app submission
might be too computationally expensive given the scale of the market,
even excluding different app versions. To test if Google Play
performs such a heavy-weight analysis, 
we released an unmodified clone of the Open Sudoku app, with
all the market signals completely modified.
This app was accepted for
publication, suggesting that the Play Store vetting process does not
perform code-level analysis. We left the app on the market 
for three months before removing it ourselves.

\parax{Experiment \#2: Metadata similarity}
Google Play's Developer Program Policy specifically prohibits
impersonation~\cite{googlePlayImpersonation}, and publishing an app with
highly similar market signals should therefore be prohibited. We released
another clone of Open Sudoku, this time with signals that were equal to
the original apps whenever possible, or highly similar when not (\ie{}
the package name and the developer name).
We used \texttt{org.moire.attribution.opensudoku} 
instead of the original \texttt{org.moire.opensudoku}
package name, and we re-used the original name of the
developer without tildes. 
The Play Store accepted the app, which shows that impersonation is still
possible despite the official policy. We left the app on the market for
three months before removing it ourselves.

\parax{Experiment \#3: Protection of high-profile apps}
We hypothesize that Google Play prioritizes the protection of popular or
high-profile apps. To test this hypothesis, we release a clone of the
Signal Private Messenger with all the attribution signals resembling the
original app as closely as possible.
The Play Store rejected our first clone
of the app for violating the impersonation policy, specifically referring
to the app's name. We note that in a subsequent submission, they reported
issues with re-using the app's icon. Compared to the previous experiments, these rejections suggest
that the Play Store is indeed prioritizing selected, high-profile apps. 
Ultimately, our cloned app seemed to have become part of a manual vetting process. Just like the original, it contained a PayPal donation link which went against the store policies of only using Google's payment system for in-app purchases. These links were attached as screenshots to our rejection, which likely did not come from an automatic process. Meanwhile, the original app removed the PayPal payment in the app.

Our results indicate that the mechanisms currently in place to enforce
publication policies on Google Play are not perfect and that
self-reported market signals  are not trustworthy for attribution purposes, 
particularly for non-popular apps.

\subsection{Multiple Signal Attribution Graphs}
\label{subsec:multiple-signal-attribution-graphs}

As shown in \S~\ref{subsec:signal-consistency-within-markets},
attribution signals tend to correlate poorly with each other
on Google Play.
In this section, we explore the soundness and validity of using a combination of
signals for attribution.
We rely on our previously defined attribution graph
(\S\ref{sec:signal-consistency}) to represent the dataset as a graph.
This large attribution graph consists of connected components, or clusters,
that we can consider to be associated with a particular author entity.
While we acknowledge that there might be a legitimate reason for one
company to have more than one value for its metadata (\eg different departments
or subsidiaries),
this reduces the attribution power of observed
signals and makes it hard to group all apps belonging to the same
owner.

The resulting graph contains \finished{158,879} clusters for
\finished{804,041} market entries.
We find that \finished{76,054} (\finished{9.5}\%) market entries are isolated
in their own cluster, and that the largest cluster
comprises \finished{288,218} (\finished{35.8\%}) market entries.
Furthermore, \finished{5,277} (\finished{3.3\%}) clusters are
\emph{fully consistent} (\ie all apps
in these non-isolated clusters are published with the same signals, except for the app
name), indicating that the majority of the clusters contain ambiguous
information.
Whereas most APKs are signed with a single certificate, we find 38 and 52
apps signed with two or three certificates, respectively.
In these cases, the app signed by multiple certificates can bridge two,
otherwise disconnected, apps with each other.

\parax{Node centrality}
The size of the one large cluster suggests that a naive attribution graph
creation can lead to substantial over-attribution. To identify the
signal causing this over-attribution, we compute the \emph{betweenness centrality} of each
signal, a metric expressing the fraction of the shortest paths in
the graph going through a particular node~\cite{10.2307/3033543}. Our intuition is
that the most central node must be key
for connecting vertices in the graph that
would otherwise not be connected, and therefore is a likely cause for over-attribution.
When looking at the centrality value of nodes, we see that the vast majority
have a centrality close to zero. There are, nonetheless, a few outliers exhibiting a relatively high
centrality.

By investigating highly-central signals, we find patterns that make
attribution challenging. App names are not unique on the Play Store,
thereby making it difficult to differentiate between apps sharing a name.
Examples include \emph{BMI Calculator} (published \finished{87} times
by \finished{86} different developer names), \emph{Flashlight}
(published \finished{153} times by \finished{149} developer names) and
\emph{Music Play} (published \finished{113} times by \finished{110} developer names).
We also observe that the highest centrality corresponds to a set
of highly prevalent signing certificates. Relying on the
subject information, we manually analyze the top 10 companies (by
centrality of the certificate) and observe that all are related to
frameworks or companies that build apps for others~\cite{oltrogge2018rise}.
The signal with the highest centrality is a certificate associated
with Andromo, which is used to sign \finished{19,096} apps. As previously
mentioned, Andromo is an app
development framework to build apps based on pre-existing
components~\cite{andromo} that are all signed by the same
certificate and are ready to submit for publication on the market.
Andromo does not mention this
signing practice in their terms of service~\cite{andromoTos}, and in fact recommends
against enrolling for signing delegation to the Play
Store~\cite{andromo-app-google-play-publication}.
Apps built with Andromo have been published under \finished{1,712} different developer
names, which---other than the signing certificate---have little in common with
each other.
In this particular case, the package name also indicates the use of this
development framework, as \finished{86.2}\% of Andromo apps follow the package name scheme
\texttt{\{com,net\}.andromo.dev<dev\_id>.app<app\_id>}. Using the same strategy of finding other highly-central certificate nodes, we
find other popular app builders, who are collectively responsible for signing
tens of thousands of apps (\eg Seattle Cloud, Bizness Apps or Mobincube sign over 7k, 3k and 1k apps respectively).

Our clustering approach reveals another app builder through the privacy policy
associated with AppsGeyser~\cite{appgeyser}, which is shared by
\finished{458} package names that otherwise share no other signal.
We also find two privacy policy URLs that serve generic privacy policies
related with products (mostly SDKs) intended 
to be used across different apps (\ie{}
Firebase~\cite{firebase-ppu} and MyAppTerms~\cite{myappterms-ppu}),
which are used by \finished{226} and \finished{390} package names, respectively.

One consequence of this practice is that customers are fully reliant on the
app builder to provide updates for the app, as the initial signing
certificate is required to do so. In addition, as we discuss in
\S\ref{sec:discussion}, certificates have use cases beyond enforcing update
integrity. As a result, the app builders have full control over the
declared and used permissions, and can automatically grant permissions
across apps from different authors without user awareness. Note that in
Android \texttt{signature permissions} are automatically granted to apps
signed with the same certificate~\cite{perms}. From an attribution
standpoint, these certificates become meaningless to identify the author
behind an app. This highlights the separation of roles that makes
attribution and accountability extremely hard on Google Play. The company
in charge of publishing---and, presumably, the one that should be accountable
for potential privacy and security issues---is not the same as the one that
has developed, or even signed, the app. In cases such as Andromo,
the developing process of the app happens automatically. This leaves a gap
between what is mandated by current legislation and market policies
and the ecosystem of Google Play apps. Specifically, 
it is typically assumed that the developer is the same as the
company publishing the app and thus the one liable for potential privacy and
security violations.

\parax{Large organizations}
The attribution graphs allows us to study the attribution signals of prominent
tech companies. We select a curated list of developer names publishing popular
apps with more than 1B installs.
The large download count serves as the ground-truth for the legitimacy of
these apps and developer accounts.
For each developer name, we collect the number
of unique signals used across apps published under the same name (\ie
account on Google Play) and measure how many other developer names use the same
certificate as the main developer account to sign their apps.
Table~\ref{tab:signals_big_orgs} shows the results of our analysis.
Most of these organizations use multiple certificates to sign their
apps with the exception of Snap, TikTok, WhatsApp, Instagram, and Skype. An extreme
case is HP, which nearly has a separate certificate per published app.
For Google, Samsung, WhatsApps and HP, we see that one or more
alternative developer name has used one of the certificates to publish an app.
Some of these names are related to the company, pointing to different units
of the company releasing the app (\eg ``Hewlett Packard Enterprise
Company'' for HP). However, as Table~\ref{tab:signals_big_orgs} shows, most
alternative names are unrelated to the original company (\eg ``Logmein, Inc''
for Samsung).
Conversely, both WhatsApp and Instagram are subsidiary companies of Facebook
Inc., but do not share any cryptographic link with their parent company.
A similar relationship holds for Microsoft and Skype.
Note that all these cases refer to companies acquired and merged
into a larger organization.
Overall, these results paint a diverse picture of publication and development strategies
across companies that can impede automatic attribution, even for software released by well-known companies.

\begin{table}[t]
\centering
    \caption{Uniqueness of signals by large organizations and the developer names
      they use with the same signing certificate.}
    \label{tab:signals_big_orgs}
    \newcommand{\mcrot}[1]{\multicolumn{1}{l}{\rlap{\rotatebox{25}{#1}~}}}
    \scalebox{1}{\setlength{\tabcolsep}{2pt}
    \begin{tabular}{p{2.5cm}rrrrp{4cm}}
        \textbf{Developer name} &
        \rot{90}{\textbf{\# apps}} &
        \hspace*{-.3em}\rot{90}{\textbf{\# emails}} &
        \hspace*{-.3em}\rot{90}{\textbf{\# websites}} &
        \hspace*{-.3em}\rot{90}{\textbf{\# certs}} &
        \textbf{Other developer names}                    \\
        \midrule
        \rowcolor{Gray}
                     Google LLC &       136 &               48 &                103 &           85 &                                                                                                                                            Google Fiber Inc.; The Infatuation Inc. \\
                       Facebook &        17 &                7 &                 13 &            7 &                                                                                                                                                                                --- \\
        \rowcolor{Gray}
        Microsoft Corporation &        83 &               62 &                 54 &           23 &                                                                                                                                                                                --- \\
 Samsung Electronics Co.,  Ltd. &        58 &               19 &                 24 &           24 &  Logmein, Inc.; Maas360; Sidi; Teamviewer; Nsl Utils; Sophos Limited; Barco Limited (Awind)\\
        \rowcolor{Gray}
        Twitter, Inc. &         2 &                2 &                  2 &            2 &                                                                                                                                                                                --- \\
                      Snap Inc &         1 &                1 &                  1 &            1 &                                                                                                                                                                                --- \\

        \rowcolor{Gray} TikTok Pte. Ltd. &         2 &                1 &                  1 &            1 &                                                                                                                                                                                --- \\
                 Netflix, Inc. &         5 &                5 &                  2 &            4 &                                                                                                                                                                                --- \\
        \rowcolor{Gray}
                  WhatsApp Inc. &         2 &                2 &                  2 &            1 &                                                                                                                                                                       Whatsapp LLC \\
                        HP Inc. &        37 &               23 &                 28 &           30 &                                                                                                                          Hewlett Packard Enterprise Company; Printeron Inc \\
        
                      \rowcolor{Gray} Instagram &         5 &                2 &                  2 &            1 &                                                                                                                                                                                --- \\
                           King &        20 &               19 &                 18 &            7 &                                                                                                                                                                                --- \\
        \rowcolor{Gray} 	    Skype &         2 &                1 &                  2 &            1 &                                                                                                                                                                                --- \\
    \end{tabular}
    }
\end{table}

\vspace{1em}
\noindent\textit{\textbf{RQ3: \rqthree}}
Our case study reveals the ubiquity of app development frameworks operating on
the Play Store. The signals that are re-used across apps created by
these frameworks (\eg certificate or privacy policy) hamper
accurate attribution. Conversely, our analysis of the publishing behaviour of
high-profile demonstrates a large variety of different signals under
which apps are published, hampering attribution too.
     \section{Concluding Remarks}
\label{sec:discussion}

Our empirical results show that automatic author attribution in the Android
ecosystem is a hard problem.
The answers to the research questions highlight the roadblocks that hamper this attribution:
\begin{itemize}[noitemsep, topsep=0pt]
    \item Not only is developer contact information sporadically provided on markets, 
	    the markets that do provide it do not
    enforce the availability of this information.
    This problem is amplified by the fact that a non-negligible fraction of apps change their market signals over time (\rqoneref).
    \item We identified frequent signal inconsistencies of apps 	
	within the publication signals of an individual app,
	across apps from a developer account on a specific market, 
	and across apps on multiple markets (\rqtworef).
    \item The publication behavior of app development frameworks 
	    and larger companies results in the association of
    individual signals with multiple developers accounts, and vice versa (\rqthreeref).
\end{itemize}

By introducing the concept of attribution graphs,
we have shown that the software developer or author
is not always the same as the company accountable for the app.
The proliferation of app development frameworks and companies that build apps
for others calls for a re-definition of roles and
for a re-adjustment of how researchers approach attribution for
different studies.

\parax{Implications}
Imprecise app attribution has negative consequences for a number of research
areas and applications as discussed in \S\ref{sec:sota}. 
It affects measurements and analyses of the
app ecosystem, market dynamics, or automatic detection of deceptive
actors and practices.
The unavailability of signals limits the software 
attribution community to rely on a combination of signals to
improve upon attribution based on single signals.
The volatility of signals threaten the validity of studies on the long term, as drawn conclusions
may not hold years later.
Our results challenge methods of attribution that prior research has resorted to.
Using unique signatures to count unique developers will undercount 
them as app-frameworks sign apps with the same certificate
for different developers 
(See \S\ref{subsec:multiple-signal-attribution-graphs})~\cite{gamba2020analysis,xucodeattribution2019,kalgutkar2018android}.
However, using the developer name or website from the market might overcount entities such as large corporations that created multiple accounts with slightly different names or domains (\S\ref{subsec:multiple-signal-attribution-graphs})~\cite{explorativemobile2017,Phishinginandroid2018,rmv2019wang}.
Our results also show that certificate-based tracking 
of apps across markets leads to
inaccurate results~(\S\ref{subsec:signal-consistency-across-markets})~\cite{lindorfer2014andradar}.

Imprecise attribution also damages transparency for users, 
as the absence of clear signals about which
company is accountable for an app limits their ability to take an informed
decision about whether to install it or not or when exercising GDPR rights. 
In the case of privacy
abuses and security vulnerabilities, it makes disclosures harder.
Furthermore, inconsistencies across markets can lead users to
install an app by mistake, only because it presents an ambiguous
signal (\eg the app name).

\parax{Signing Certificates}
We found that the assumption that certificates can be used
for attribution (\S~\ref{sec:sota}) is flawed, as one author does not
necessarily use a single certificate and a single certificate
can be used to sign apps from different organizations. The adoption of
Google's signing
delegation process by developers also reduces the value of 
signing certificates as attribution signals.
The unreliability of the signing certificate has profound
consequences for the platform security, as it has long been used as a proof of
authorship in prior work and even for threat intelligence.
The installation of two apps from different organizations signed with the same
certificate also has privacy implications: Android automatically grants permissions requested by one app to the
other app~\cite{sharedUID, gamba2020analysis}. Moreover,
certificates are also used to link apps to websites~\cite{assetlinks}, thus enabling the
website to check for \textit{``their''} installed apps~\cite{getInstallRelated} and even
share credentials without direct consent~\cite{androidsmartlock}.
These implications impact not only unaware end users, but also developers that choose to hand over the certificate signing
to another entity.

\parax{Recommendations}
Only profound platform and market changes can solve the predicament of Android attribution.
First, the Android
ecosystem should move away from self-signing certificates and
vet the information disclosed in apps' market profiles.
Other prominent software ecosystems already rely on more sound signature
mechanisms, such as Apple issuing valid certificates for its app
store~\cite{apple-dev-enroll} and Windows relying on
a PKI~\cite{kim2017certified}.
While not perfect~\cite{kotzias2015certified}, we argue that these approaches
limit the number of certificates with incomplete or invalid
information while also raising the bar for malicious actors.
The open nature of the Android ecosystem should be preserved (\eg free and automatic
certificate issuance) and should allow for the current business practises of developer to
remain in place (\eg{} certificate reuse across publishing developers and multiple certificate
use by developers).
In practise, however, 
the certificate issuance process would enforce (the very least) for certificate-related
fields to be available, but could also include verification on the values of these fields.
Showing this information on the market listing of an app on markets would also expose this
currently unavailable information to the end user.
An alternative option would be to rely
on independent third-party verification for attribution, but we argue that this would simply
shift responsibility from the market to third-party entities.
An implementation could be developed without requiring changes to the current ecosystem
and could range from automatic authorship attribution to producing warning for conflicting
signals to users.

Our work also has implications for large app repositories compiled for
research purposes. Placing app metadata in existing repositories such
as AndroZoo and antivirus
engines like VirusTotal would benefit the research community and provide
intelligence to contextualize observed software behaviors.
Prominent markets like
Google Play should include
veracity checks for developer metadata included in Google Play Protect,
while ensuring the identity of the developer as iOS does.

\parax{Future work}
Current legislation~\cite{gdpr} mandates software that collects and shares user-data
to provide a comprehensive privacy policy that should
include, among other information, data about the company behind
the software~\cite{eudirective}.
Therefore, we argue that automatically parsing privacy policies can
yield information useful for attribution. However,
automatic parsing of legal texts is a complex problem, specially
at scale~\cite{zimmeck2016automated,harkous2018polisis,andow2019policylint}.
In a preliminary analysis, we find that it is not common for
policies to contain
contact information as only 59\% and and 12\% of them have an email and
an address respectively.
Similarly, other more complex and expensive methods
such as UI inspection and code analysis
could be used for extracting attribution data~\cite{xucodeattribution2019}.
Finally, we believe that the generalizable concept of
attribution graphs has
potential applications beyond attribution in the
Android ecosystem, including large-scale market analysis, study
of developer trends and practices, and clone detection.

\section*{Acknowledgments}
This research has been partially funded by the Spanish Government grant
ODIO (PID2019-111429RB-C21 and PID2019-111429RBC22); the Region of Madrid,
co-financed by European Structural Funds
ESF and FEDER Funds, grant CYNAMON-CM (P2018/TCS-4566); and by the EU H2020
grant TRUST aWARE (101021377). This research also received funding from the Vienna Science and Technology Fund (WWTF) through project ICT19-056 (IoTIO), and SBA Research (SBA-K1), a COMET Centre within the framework of COMET – Competence Centers for Excellent Technologies Programme and funded by BMK, BMDW, and the federal state of Vienna. The COMET Programme is managed by FFG.
The opinions, findings, and conclusions or
recommendations expressed are those of the authors and do not necessarily
reflect those of any of the funders.

    \bibliographystyle{IEEEtran}


    \begin{IEEEbiography}[{\includegraphics[width=0.8in,height=1in,clip,keepaspectratio]{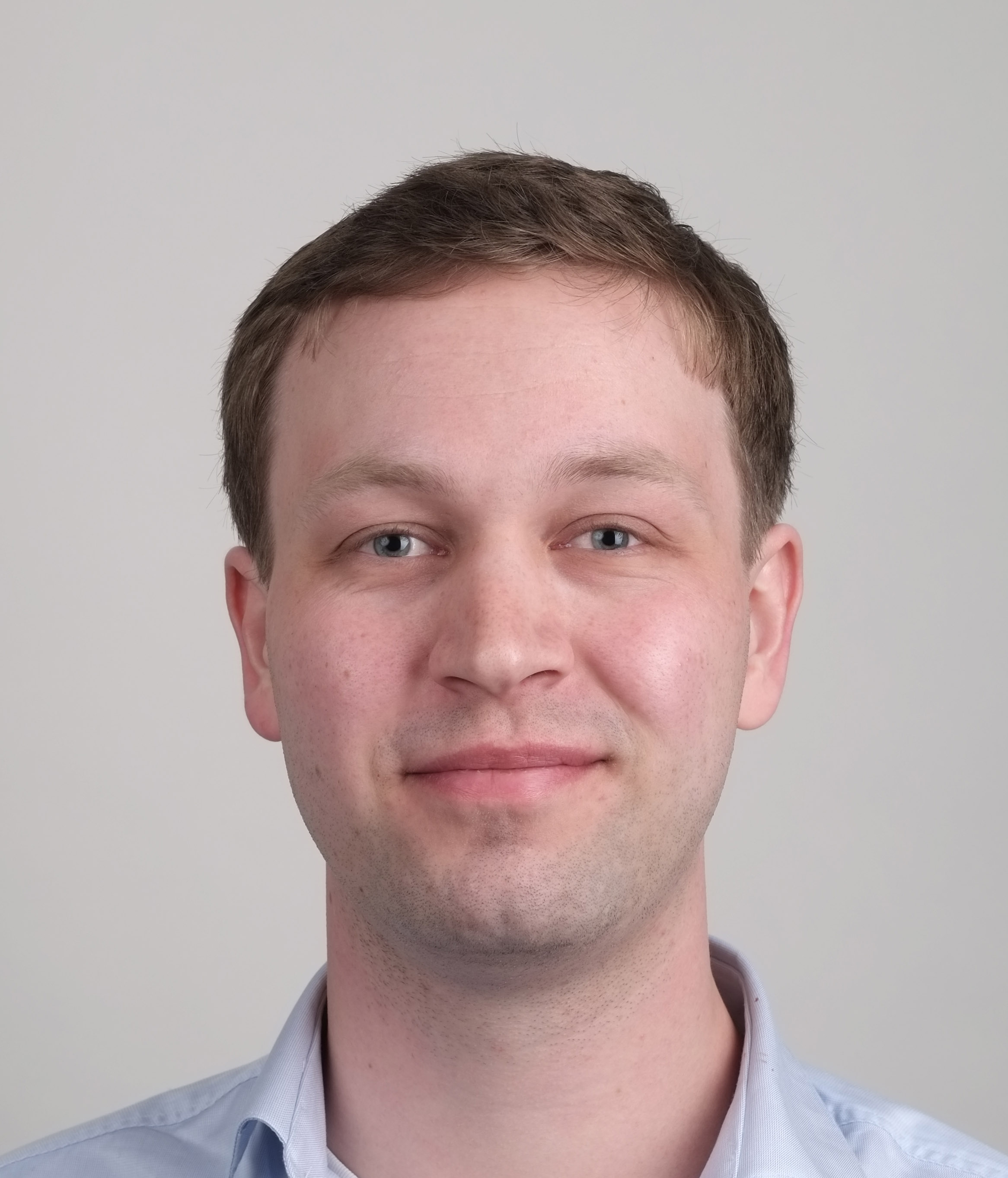}}]{Kaspar Hageman}
        is a postdoctoral researcher at Aarhus University, Denmark.
        He obtained his PhD degree at Aalborg University in 2021 on the analysis of the domain name system, digital certificates and
        network traffic for security.
        His research interests further include applying machine learning to network security and more recently collaborative drone fleets.
        His work was awarded the IETF/IRTF Applied Networking Research Award in 2017.
    \end{IEEEbiography}
    \begin{IEEEbiography}[{\includegraphics[width=0.8in,height=1in,clip,keepaspectratio]{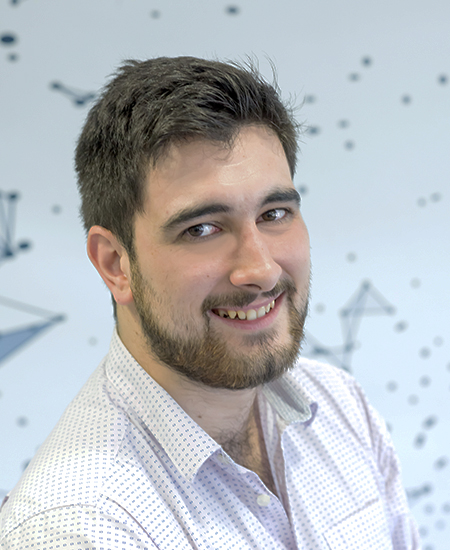}}]{\'Alvaro Feal}
        is a PhD student working at IMDEA Networks Institute under Prof.
        Narseo Vallina-Rodriguez's supervision.
        His research revolves around analyzing privacy threats in the mobile and web
        ecosystem using static and dynamic analysis techniques as well as network
        measurements. He has published his research in different venues such as
        the IEEE Symposium on Security and Privacy, USENIX Security, ACM IMC, PETS
        Symposium, IEEE ConPro, and CPDP. He has received several awards such as
        the Distinguished Paper Award at Usenix Security’19 and the
        ``Premio de Investigación en Datos Personales: Emilio Aced'' in 2021
        and 2022.

    \end{IEEEbiography}
    \begin{IEEEbiography}[{\includegraphics[width=0.8in,height=1in,clip,keepaspectratio]{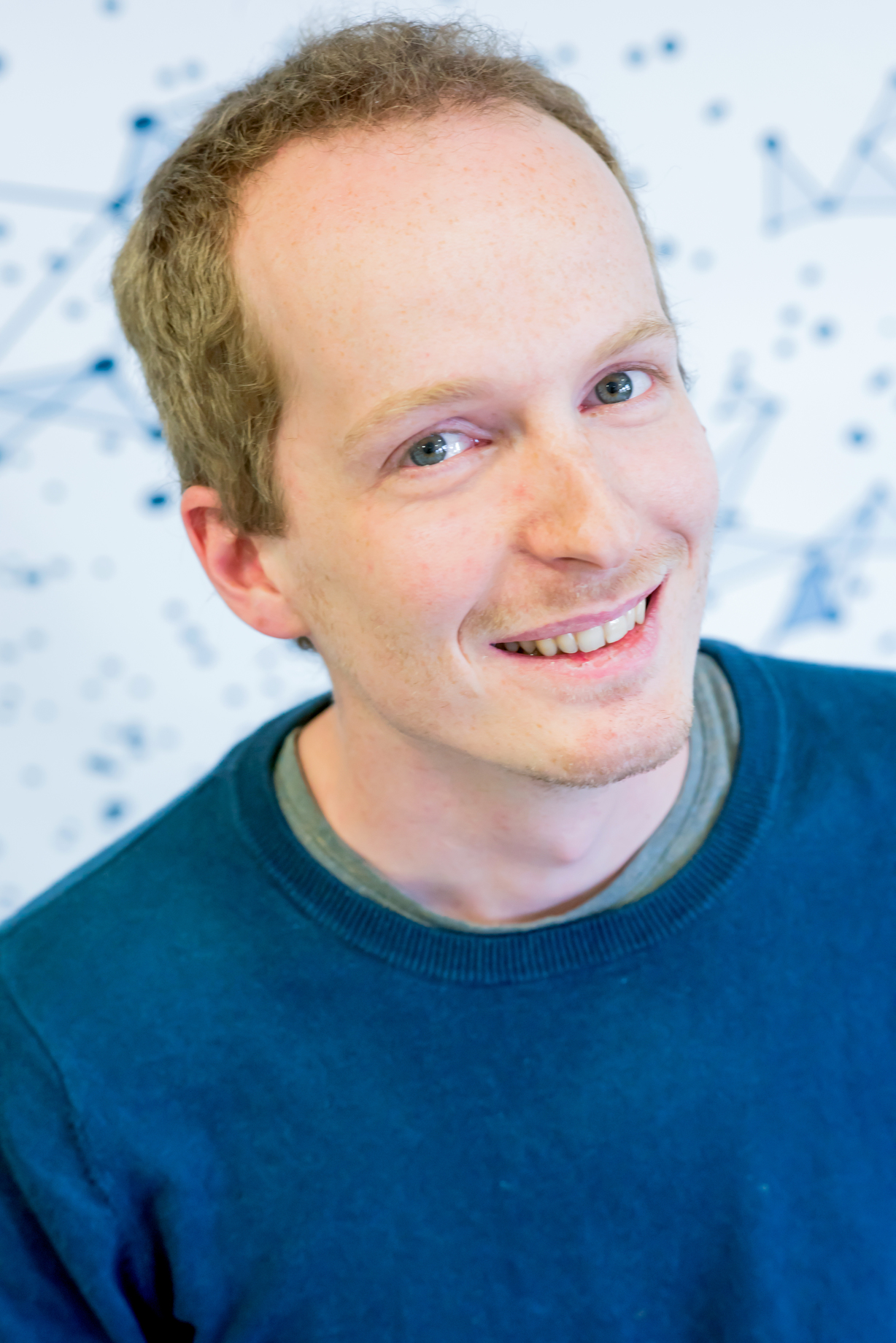}}]{Julien Gamba}
        is a PhD student in the Internet Analytics Group at the IMDEA Networks
        Institute.  His research revolves around user's security and privacy in
        Android devices. In his work, Julien uses both static and dynamic
        analysis, as well as other techniques specifically designed to
        understand the behavior of mobile applications. Recently, Julien was
        the first author of the first large-scale analysis of the privacy and
        security risks of pre-installed software on Android devices and their
        supply chain, which was awarded the Best Practical Paper Award at the
        41st IEEE Symposium on Security and Privacy.  This study was featured
        in major newspaper such as The Guardian (UK), the New York Times (USA),
        CDNet (USA) or El País (Spain).  Julien was also awarded the ACM IMC
        Community Contribution Award in 2018 for his analysis of domain ranking
        services, and was awarded the NortonLifeLock Research Group Graduate
        Fellowship, the Google PhD Fellowship in Security and Privacy and
        Consumer Reports' Digital Lab fellowship.
    \end{IEEEbiography}
    \begin{IEEEbiography}[{\includegraphics[width=0.8in,height=1in,clip,keepaspectratio]{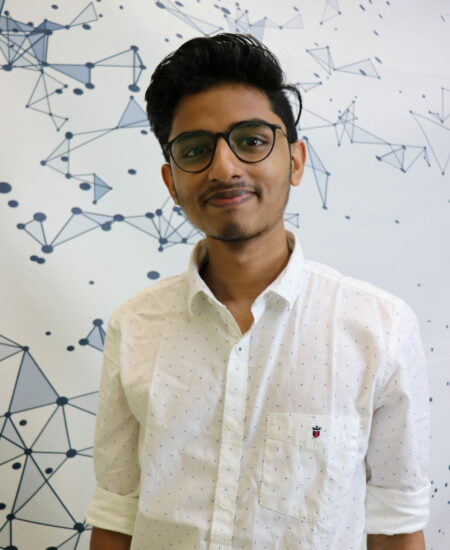}}]{Aniketh
        Girish} is PhD student working at the Internet Analytics group of IMDEA
        Networks Institute and Universidad Carlos III de Madrid advised by Prof.
        Narseo Vallina-Rodriguez. Aniketh empirically measure the privacy and
        security risks of smart home and Mobile ecosystem with a particular
        focus on consumer data protection and regulatory compliance. Aniketh has
        published his research in international peer-reviewed conferences such
        as USENIX Security’20.
    \end{IEEEbiography}
    \begin{IEEEbiography}[{\includegraphics[width=0.8in,height=1in,clip,keepaspectratio]{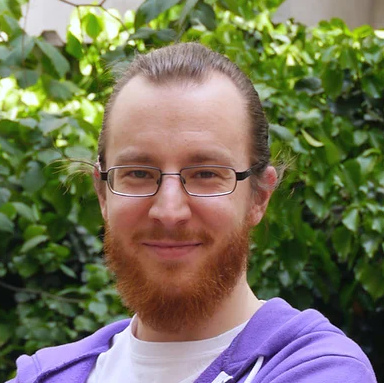}}]{Jakob Bleier}
        is a PhD student at the TU Wien Security and Privacy Research Unit under 
        the supervision of Martina Lindorfer. He is working in the area of system 
        security and is developing new methods and tools for measuring 
        similarities between programs, as well as identifying the libraries used.
        His work has contributed to a publication presented at the AsiaCCS’22 conference.
    \end{IEEEbiography}
    \begin{IEEEbiography}[{\includegraphics[width=0.8in,height=1in,clip,keepaspectratio]{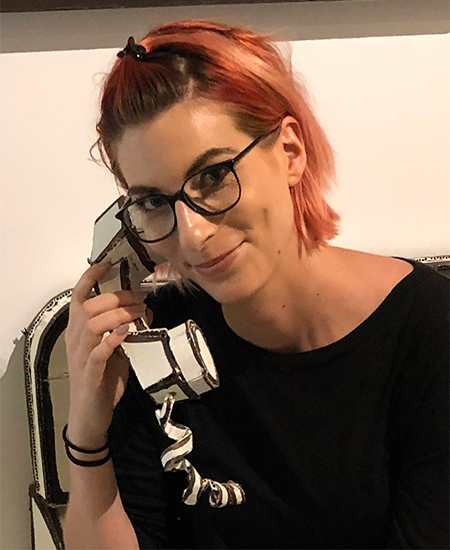}}]{Martina Lindorfer} is an Assistant Professor at TU Wien, Austria, and a key researcher at SBA Research, the largest research center in Austria which exclusively addresses information security. Prior to starting her tenure track she spent two years as a postdoc at the University of California, Santa Barbara. Her research focuses on applied systems security and privacy, with a special interest in automated static and dynamic analysis techniques for the large-scale analysis of applications for malicious behavior, security vulnerabilities, and privacy leaks. 
        Her research and outreach activities have been recognized with the ERCIM Cor Baayen Young Researcher Award, the ACM Early Career Award for Women in Cybersecurity Research, as well as the Hedy Lamarr Award from the City of Vienna. 
    \end{IEEEbiography}
    \begin{IEEEbiography}[{\includegraphics[width=0.8in,height=1in,clip,keepaspectratio]{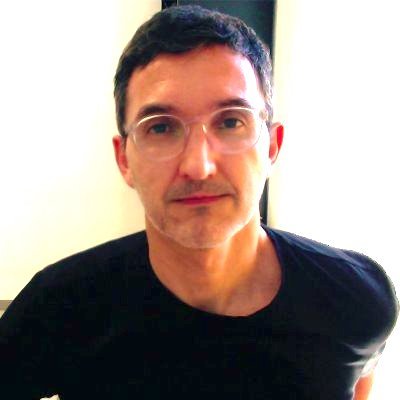}}]{Juan Tapiador}
         is Professor of Computer Science at Universidad Carlos III de
         Madrid, Spain, where he leads the Computer Security Lab. Prior to
         joining UC3M he worked at the University of York, UK. His research
         interests include systems security, malware, privacy, and network
         measurements. He has served in the technical committee of
         conferences such as USENIX Security, ACSAC, DIMVA, ESORICS and
         AsiaCCS. He has been the recipient of the UC3M Early Career Award
         for Excellence in Research (2013), the Best Practical Paper Award
         at the 41st IEEE Symposium on Security and Privacy (Oakland), the
         CNIL-Inria 2019 Privacy Protection Prize, and the 2019 AEPD Emilio
         Aced Prize for Privacy Research. His work has been covered by
         international media, including The Times, Wired, Le Figaro, and
         The Register.
    \end{IEEEbiography}
    \begin{IEEEbiography}[{\includegraphics[width=0.8in,height=1in,clip,keepaspectratio]{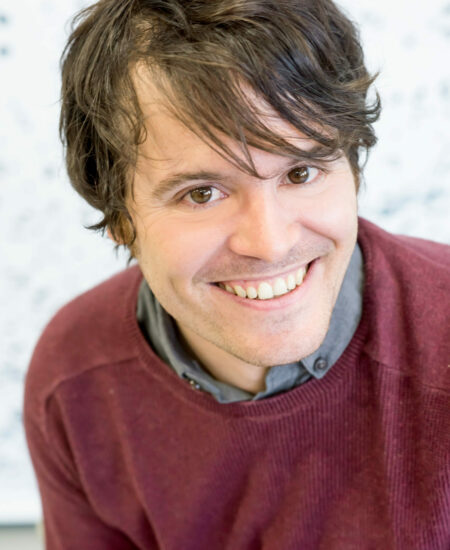}}]{Narseo Vallina-Rodriguez}
        is an Associate Research Professor at
        IMDEA Networks and a co-founder of AppCensus Inc.
        Narseo obtained his Ph.D. at the University of Cambridge and
        his research interests fall in the broad areas
        of network measurements, privacy, and mobile security.
        His research efforts have been awarded with best paper awards at the 2020 IEEE
        Symposium on Security and Privacy (S\&P), USENIX Security’19, ACM
        IMC’18, ACM HotMiddlebox’15, and ACM CoNEXT'14 and Narseo has
        received prestigious industry grants and awards such as a Google Faculty Research
        Fellowship, a DataTransparencyLab Grant, and a Qualcomm Innovation Fellowship.
        His research in the mobile security and privacy domain has been covered by
        international media outlets like The Washington Post, The New York
        Times, and The Guardian and it has influenced policy changes and security improvements in the Android
        platform.
        Narseo's work has received in multiple occasions the recognition of EU Data Protection Agencies with
        the AEPD Emilio Aced Award (2019, 2020, and 2021) and the CNIL-INRIA Privacy Protection
        Award (2019 and 2021).
        He is also the recipient of the IETF/IRTF Applied Networking Research Award
        in 2016 and the Caspar Bowden Award in 2020.
    \end{IEEEbiography}

\end{document}